\documentclass[fleqn,usenatbib]{mnras}

\usepackage{mathptmx}
\usepackage{xcolor}
\usepackage{natbib}
\usepackage[T1]{fontenc}

\DeclareRobustCommand{\VAN}[3]{#2}
\let\VANthebibliography\thebibliography
\def\thebibliography{\DeclareRobustCommand{\VAN}[3]{##3}\VANthebibliography}

\usepackage{graphicx}	
\usepackage{amsmath}	
\usepackage{amssymb}	

\title{Forward modelling of \textit{Kepler}-band variability due to faculae and spots}

\author[L. J. Johnson et al.]{
Luke J. Johnson,$^{1}$\thanks{E-mail: l.johnson17@imperial.ac.uk}
Charlotte M. Norris,$^{1}$
Yvonne C. Unruh,$^{1}$
Sami K. Solanki,$^{2,3}$
\newauthor Natalie Krivova,$^{2}$
Veronika Witzke$^{2}$
and Alexander I. Shapiro$^{2}$
\\
$^{1}$Department of Physics, Imperial College London, London SW7 2AZ, UK\\
$^{2}$Max Planck Institute for Solar System Research, Justus-von-Liebig-Weg 3, 37077 G{\"o}ttingen, Germany\\
$^{3}$School of Space Research, Kyung Hee University, Yougin, 446-701, Gyeonggi, Republic of Korea
}

\date{Accepted 2021 April 22. Received 2021 April 21; in original form 2021 February 24}

\pubyear{2021}

\begin{document}
\label{firstpage}
\pagerange{\pageref{firstpage}--\pageref{lastpage}}
\maketitle

\begin{abstract}
Variability observed in photometric lightcurves of late-type stars (on timescales longer than a day) is a dominant noise source in exoplanet surveys and results predominantly from surface manifestations of stellar magnetic activity, namely faculae and spots. The implementation of faculae in lightcurve models is an open problem, with scaling typically based on spectra equivalent to hot stellar atmospheres or assuming a solar-derived facular contrast. We modelled rotational (single period) lightcurves of active G2, K0, M0 and M2 stars, with Sun-like surface distributions and realistic limb-dependent contrasts for faculae and spots. The sensitivity of lightcurve variability to changes in model parameters such as stellar inclination, feature area coverage, spot temperature, facular region magnetic flux density and active band latitudes is explored. For our lightcurve modelling approach we used \texttt{actress}, a geometrically accurate model for stellar variability. \texttt{actress} generates 2-sphere maps representing stellar surfaces and populates them with user-prescribed spot and facular region distributions. From this, lightcurves can be calculated at any inclination. Quiet star limb darkening and limb-dependent facular contrasts were derived from MURaM 3D magnetoconvection simulations using ATLAS9. 1D stellar atmosphere models were used for the spot contrasts. We applied \texttt{actress} in Monte-Carlo simulations, calculating lightcurve variability amplitudes in the \textit{Kepler} band. We found that, for a given spectral type and stellar inclination, spot temperature and spot area coverage have the largest effect on variability of all simulation parameters. For a spot coverage of $1\%$, the typical variability of a solar-type star is around $2$ parts-per-thousand. The presence of faculae clearly affects the mean brightness and lightcurve shape, but has relatively little influence on the variability.
\end{abstract}





\section{Introduction} \label{intro}

Over its 9.5 year operation, the NASA \textit{Kepler} satellite \citep{Borucki2010} collected high precision photometry on $\sim 5 \times 10^5$ stars, resulting in the confirmed detection of $>2500$ exoplanets. Activity-induced stellar variability is a dominant source of the intrinsic noise in both transit (\citealp{Czesla2009}, \citealp{Oshagh2014} and \citealp{Kirk2016}) and radial velocity surveys (\citealp{Saar1997}, \citealp{Desort2007}, \citealp{Lagrange2010}, \citealp{Meunier2010a}, \citealp{Haywood2014b} and \citealp{Oshagh2017}) that can hinder the accurate detection and characterisation of planets with active stellar hosts. As a result, \textit{Kepler} has detected fewer low-mass planets than expected (\citealp{Korhonen2015} and \citealp{Andersen2015}). In the process however, the high-precision photometry collected has provided information on stellar photometric variability \citep{Basri2013}, which will continue to be an important consideration when analysing data from \textit{Kepler} successors such as \textit{TESS}, \textit{CHEOPS}, \textit{PLATO} and \textit{ARIEL}.

Lightcurves of late-type stars show a quasi-periodic behaviour on timescales relevant for planetary transits (hours to days). This variability is mainly caused by the presence of photospheric surface features, manifestations of surface magnetic activity, which evolve in time and appear and disappear on a star's visible hemisphere as it rotates. The most prominent of these features are dark spots and bright faculae (\citealp{Berdyugina2004}, \citealp{Domingo2009}, \citealp{Solanki2013} and \citealp{Radick2018}).

The link between the presence of surface features and stellar variability can be uncovered by observing the Sun as a star (\citealp{Wilson1991} and \citealp{Frohlich2013}). By comparing high-resolution solar disc images and simultaneous solar photometry, spot and facular signatures can be identified in solar lightcurves. Large dips are observed in solar lightcurves at times corresponding to spot crossings, and double-peaked signatures are observed during the passage of faculae across the solar disc (\citealp{Fligge1998}, \citealp{Fligge2000} and \citealp{Frohlich2002}). The double-peaked shape results from solar faculae having large intensity contrasts near the limb of the disc, but being difficult to distinguish from the quiet photosphere close to disc centre in the visible broadband spectrum.

For more distant stars, photometric lightcurve analysis is often used to infer the presence and surface area coverage of surface features. However, there are limitations to this approach: photometry alone cannot provide reliable information on feature latitudes and is prone to underestimating coverage areas as it is insensitive to rotationally invariant features such as latitudinal bands and polar features \citep[see][and references therein]{Solanki2004}. This can be overcome by taking measurements in more than one band, allowing colour excesses to be detected and pointing to the presence of rotationally invariant features. There also exists a degeneracy between spot areas and contrasts (\citealp{Wolter2009} and \citealp{Silva-Valio2010b}). This is because a larger lightcurve dip can result from either a spot with larger area or lower brightness/temperature (although the width can help constrain the feature size). \citet{Basri2018} noted that almost all stellar lightcurves can be fitted using a `2-spot' model (see also \citealp{Torres1972}, \citealp{VanLeeuwen1987} and \citealp{Rodrigo2021a}). The fact that we never see `simple' feature distributions like this on the Sun raises the question of whether accurate spatial information on stellar surface feature distributions can be inferred through lightcurve analysis.

In order to obtain a metric of a star's activity level from photometric observations, \citet{Basri2011} introduce the range variability, a simple, but useful quantity computed as $R_{\rm var} = P(95) - P(5)$, where $P(5)$ and $P(95)$ are the $5^{\rm th}$ and $95^{\rm th}$ percentiles of the lightcurve amplitude respectively (calculated over a chosen time interval). In \citet{Basri2013}, $R_{\rm var}$ was calculated for the Sun during Solar Cycle 23 over $30$ day intervals (just over one solar rotation period, $P_{\odot} \approx 24.5\,{\rm days}$ at the equator). $R_{\rm var}$ correlates very well with other activity indices on the Sun \citep{Salabert2017} and reflects the solar cycle. Stars with larger amounts of surface features are generally expected to have higher $R_{\rm var}$ values. While $R_{\rm var}$ is sensitive to instrumental noise when at comparable levels to that of the stellar activity or higher, it is much more robust than the untrimmed lightcurve amplitude (see Appendix \ref{noise}).


Several recent studies have focused on forward modelling lightcurves of active late-type stars, e.g. \citet{Rackham2018, Rackham2019} and \citet{Sarkar2020}. These studies use synthetic spectra derived from 1D model atmospheres to approximate intensities for the quiet star and surface features at chosen effective temperatures. While this approach is sufficient to model spots, it neglects the strong dependence of facular intensities on their position on the stellar disc, treating them as `hot spots'. At most wavelengths, solar faculae appear brighter when viewed at a greater angle near the disc limb because the magnetic flux tubes that manifest as faculae at the photosphere radiate mostly through their hot walls \citep{Spruit1976}. In \citet{Herrero2016} and \citet{Cauley2017}, the empirically-constructed facular limb brightening law of \citet{Meunier2010b} is used to reproduce this effect. This was extrapolated to a solar model \citep{Borgniet2015} and later to stars of spectral type F6-K4 (\citealp{Meunier2019a, Meunier2019b}) based on preliminary runs of the simulations used in this paper \citep{Norris2018}.


Since there are no direct observations of faculae on stars other than the Sun, the only possible solution is to rely on simulations that are capable of taking the interaction between the matter and magnetic field into account, i.e. magnetohydrodynamics (MHD) simulations (e.g. \citealp{Salhab2018}). 3D magnetoconvection simulations were used to calculate both quiet star (\citealp{Danilovic2010}, \citealp{Hirzberger2010} and \citealp{Beeck2013a}) and facular (\citealp{Afram2011} and \citealp{Beeck2015b}) intensity contrasts for a range of stellar spectral types. In \citet{Beeck2013a, Beeck2015b}, the 3D radiation-MHD code MURaM \citep{Vogler2005} was used to simulate 3D regions of a solar twin atmosphere for a range of surface field strengths. \citet{Norris2017} used the ATLAS9 spectral synthesis code \citep{Castelli2003} emergent intensities from these regions. Consequently, mean intensity contrasts were derived for the quiet photosphere and facular regions with a range of mean magnetic field strengths in the wavelength range $149.5\,{\rm nm} \leq \lambda \leq 160000.0\,{\rm nm}$, finding good agreement with solar observations at different heliocentric angles. Facular contrast values based on MURaM calculations were used in \citet{Schrijver2020} to model photospheric facular appearances on Sun-like stars in the ultraviolet (UV), visible and infrared (IR).

We developed the stellar variability code \texttt{actress} to use quiet star intensities and facular contrasts derived from MURaM simulations \citep{Norris2017} in forward modelling the lightcurves of active G2, K0, M0 and M2 stars. We first determine the lightcurve signatures resulting from individual surface features for different spectral types and viewing angles in the \textit{Kepler} band. Next, we use a Monte-Carlo approach, populating \texttt{actress} model stars with Sun-like feature distributions and calculating rotational lightcurves and their corresponding $R_{\rm var}$ to quantify the possible effect of varying our physical parameter inputs on variability levels. 

The paper is structured as follows: in Sect.~\ref{modelling}, our lightcurve modelling approach is detailed. The simulation inputs and setup are described in Sect.~\ref{distributions}. Results of the investigations performed are presented in Sect.~\ref{results} and discussed in Sect.~\ref{disc}. The conclusions are given in Sect.~\ref{conc}.

\section{Modelling stellar lightcurves} \label{modelling}

\begin{figure}
\centering
\includegraphics[width=\hsize]{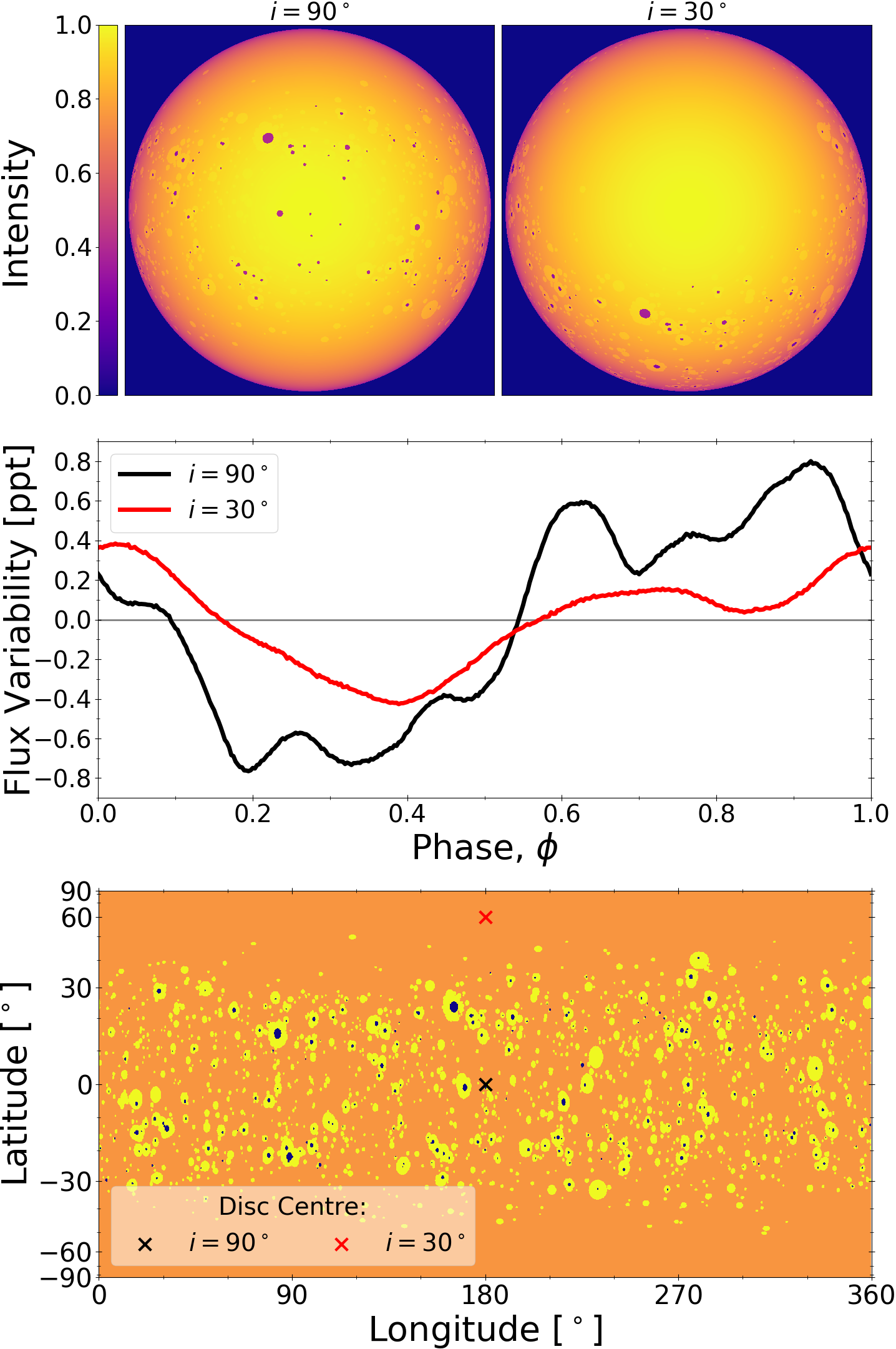}
\caption{\label{fig:actsol} Example of \texttt{actress} lightcurve modelling. Top panel: normalised intensity maps of a limb darkened, solar-type star viewed in the \textit{Kepler} band at rotational phase $\phi = 0.5$ with stellar inclinations $i=90^\circ$ (left) and $30^\circ$ (right). Middle panel: Corresponding lightcurves calculated with \texttt{actress} at inclinations $i=90^\circ$ (black line) and $30^\circ$ (red line). Bottom panel: HealPix map representing the active stellar surface, cosine-scaled in latitude and flattened in longitude to resemble a solar synoptic map. The quiet photosphere is displayed in orange, facular regions are bright yellow and spot regions are dark blue. The crosses represent the centres of the stellar discs in the top panel. We note that the brightness of the features on the HealPix map does not correspond to the actual brightness (which is a function of observing angle).
}
\end{figure}

\texttt{actress} calculates model lightcurves for rotating, magnetically active stars. The active stellar surface is modelled using a spherical HealPix map and populated with circular spots and facular regions with defined radii and positions (see Sect.~\ref{distributions}). HealPix allows 2D orthographic hemispherical projections to be taken from any angle, representing the stellar disc as viewed by an observer. Limb-dependent intensities are assigned to each pixel for the quiet photosphere and surface features (see Sect.~\ref{intcalc}). Lightcurves are calculated by rotating the star at a chosen inclination $i$ (the angle between the line of sight and the rotation axis, $i=90^\circ$ corresponds to an equator-on view of the star) and accounting for the contributions of all pixels on the visible disc throughout a full rotational phase.

An example of the lightcurve modelling approach is shown in Fig.~\ref{fig:actsol} for a Sun-like star. In the top panel, intensity maps of the stellar disc are shown, with stellar inclinations $i=90^\circ$ and $30^\circ$. Dark spots are visible regardless of disc position (or viewing angle) but bright facular regions are more prominent towards the disc limbs. The resulting lightcurves shown in the middle panel reflect the surface distribution of features throughout a rotation. As the features are mostly positioned close to the equator in latitude, fewer surface features are visible at $i=30^\circ$ than at $i=90^\circ$. The $i=30^\circ$ lightcurve is both smoother and has a smaller amplitude as a result (see Fig.~\ref{fig:lcso} for equivalent lightcurves calculated with spots only). The HealPix map of the model photosphere in the bottom panel shows that the largest lightcurve dips correspond to the longitude regions most densely populated with spots. Numerical noise is present in the simulated lightcurves, resulting from the pixellation of features on both the surface map and disc projection. No additional noise is present.

\subsection{Intensity calculations} \label{intcalc}

To model active stellar discs, emergent intensities are needed as a function of disc position, or limb distance $\mu = \cos{\theta}$ where $\theta$ is the viewing angle.  Limb-dependent intensities for the quiet photosphere and surface features are obtained by calculating local thermodynamic equilibrium (LTE) specific intensity spectra with ATLAS9 (\citealp{Castelli2003} and \citealp{Kurucz2017}) at 9 viewing angles, corresponding to limb distances $0.2 \leq \mu \leq 1.0$ (\citealp{Norris2017} and \citealp{Norris2018}). Figure~\ref{fig:hydroint} (inset) illustrates how $\mu$ corresponds to disc position. The intensity spectra for the quiet photosphere and facular regions are calculated along sight lines through MURaM simulation cubes at different viewing angles, taking 3D geometry into account. Intensity spectra for the spots are calculated using 1D model atmospheres instead. This approach has been well-tested on the Sun and shown to reproduce total solar irradiance with very high accuracy \citep{Yeo2017}.

In a given wavelength band (the \textit{Kepler} band was used in this study), we fit a 3-parameter non-linear limb darkening relation,
\begin{equation}
I(\mu) = I(1) \left( 1 - a(1-\mu) - b(1-\mu^{\frac{3}{2}}) - c(1-\mu^2) \right),
\label{eq:limb_darkening}
\end{equation}
to the 9 angles from computations, following \citet{Sing2009} (see also \citealp{Sing2010}). $I(1)$ is the intensity at disc centre and $a$, $b$ and $c$ are limb darkening coefficients. This allows us to obtain the intensity at any limb distance. The fit residuals are very small for all intensity profiles, with an average fractional difference of $0.5\,{\rm ppt}$ from the computed intensity values (and a maximal difference of $2.2\,{\rm ppt}$).



\subsection{MURaM-derived intensities for quiet stars and faculae} \label{muramint}

\begin{table}
\caption{Limb-dependent \textit{Kepler}-band intensity coefficients for the 3-parameter non-linear limb darkening law (Equation \ref{eq:limb_darkening}), for all available spectral types and magnetic field strengths in our modelling approach.}      
\label{table:LD} 
\centering             
\begin{tabular}{c c c c c c}  
\hline\hline                 
Type  & $\langle B_{\rm z} \rangle$ & $I$(1) & $a$ & $b$ & $c$ \cr & [G] & [photons/$\rm m^{2}$/s/sr] & & &\\   

\hline                        
      &  hydro & $2.34 \times 10^{21}$ & 1.51 & -1.21 & 0.40\\
  G2  &  100 & $2.37 \times 10^{21}$ & 1.62 & -1.40 & 0.48\\
      &  500 & $2.33 \times 10^{21}$ & 2.33 & -2.77 & 1.09\\
      \hline
      &  hydro & $1.10 \times 10^{21}$ & 0.93 & -0.14 & -0.06\\
  K0  &  100 & $1.12 \times 10^{21}$ & 1.08 & -0.43 & 0.07\\
      &  500 & $1.11 \times 10^{21}$ & 1.30 & -0.87 & 0.27\\
      \hline
      &  hydro & $2.80 \times 10^{20}$ & 1.83 & -1.62 & 0.54\\
  M0  &  100 & $2.81 \times 10^{20}$ & 1.85 & -1.66 & 0.55\\
      &  500 & $2.77 \times 10^{20}$ & 1.77 & -1.60 & 0.53\\
      \hline
      &  hydro & $1.86 \times 10^{20}$ & 1.69 & -1.39 & 0.42\\
  M2  &  100 & $1.86 \times 10^{20}$ & 1.61 & -1.23 & 0.33\\
      &  500 & $1.80 \times 10^{20}$ & 1.86 & -1.80 & 0.65\\
\hline                                   
\end{tabular}
\end{table}

Solar faculae form when small magnetic flux tubes intersecting the stellar surface are swept by convective motion into intergranular lanes and the field is there intensified by the process of convective collapse (\citealp{Parker1978}, \citealp{Spruit1979} and \citealp{GrossmannDoerth1998}; see \citealp{Solanki1993} for a review). Plasma is evacuated from within the flux tubes, decreasing the opacity. The decrease of the convective heating is overcompensated by the radiative heating from hot walls, resulting in regions brighter than the surrounding photosphere \citep{Spruit1976}. At wavelengths dominated by continuum radiation in the visible, strong faculae appear brightest when viewed side on (near the disc limb) as the hot walls are most visible (\citealp{Yeo2013}). An approach that accurately accounts for 3D geometry is required to reproduce this behaviour (\citealp{Keller2004} and \citealp{Carlsson2004}). Simulations are the only way to obtain reliable facular contrasts for stars of spectral types other than G2.

\begin{figure}
\centering
\includegraphics[width=\hsize]{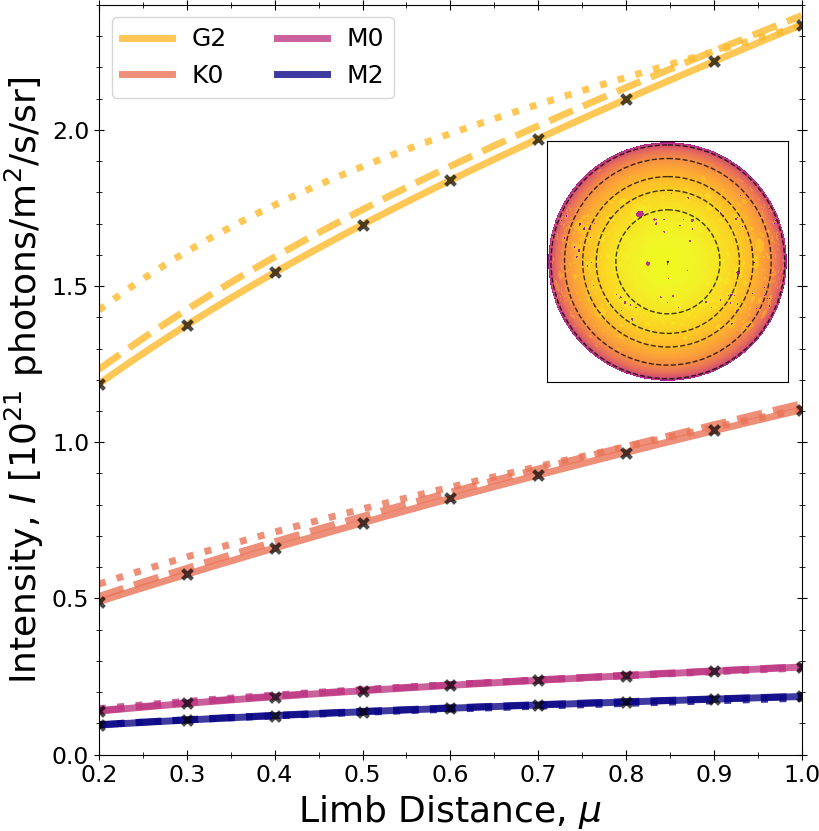}
\caption{\label{fig:hydroint}Plot of MURaM-derived average intensity fits for the quiet photosphere (solid lines), weak-field facular regions (dashed lines) and strong-field facular regions (dotted lines) against limb distance in the \textit{Kepler} band for G2, K0, M0 and M2 stars. Black crosses represent the 9 computed quiet star intensity values for each spectral type. Inset: \texttt{actress}-generated G2 stellar disc with lines of constant $\mu$. The central point represents $\mu = 1.0$, and the dashed lines (moving radially outwards from the centre) represent $\mu = 0.9$, $0.8$, $0.7$, $0.5$ and $0.2$.}
\end{figure}

A stellar disc without magnetic field is devoid of bright and dark surface features with lifetimes of days or longer\footnote{The constantly evolving cellular pattern on the solar surface is made up of bright granules and dark intergranular lanes \citep{Leighton1963}. These small-scale brightness variations occur on much shorter timescales ($\lessapprox 5$ hours) \citep{Shapiro2017}.} and will appear limb darkened. For the quiet photosphere, the MURaM code (\citealp{Vogler2005} and \citealp{Beeck2013a}) was used to run magnetoconvection simulations without magnetic fields (referred to as hydrodynamic, or field-free). Once relaxed, several simulation snapshots were taken at constant time intervals. To obtain quiet star intensities for all spectral types in our modelling approach, spectra were calculated along rays through the simulation cube snapshots at 9 viewing angles and averaged to obtain mean spectra at all angles, following \citet{Norris2017}. We then convolve the spectra with the transmission curve of a given passband and integrate to obtain intensities (in photon units). Equation \ref{eq:limb_darkening} was fitted with the \textit{Kepler}-band intensities. The resulting quiet star intensity relations are shown in Fig.~\ref{fig:hydroint} (solid lines), for all 4 spectral types considered in our models. We provide \textit{Kepler}-band quiet star fit coefficients in Table.~\ref{table:LD}, labelled `hydro' (see Appendix \ref{app:tessfit} for quadratic fit coefficients and equivalent \textit{TESS}-band coefficients).

For the facular regions, vertical fields of mean strength $\langle B_{\rm z} \rangle = 100\,{\rm G}$ and $500\,{\rm G}$ were injected into the field-free MURaM simulations\footnote{The intrinsic field strengths of individual magnetic features in the averaged $100\,{\rm G}$ and $500\,{\rm G}$ runs are roughly the same, but the runs are referred to as such for simplicity.}. Once again, the simulations were allowed to relax before a series of snapshots were taken and mean spectra calculated. The facular region intensities derived from the $100\,{\rm G}$ and $500\,{\rm G}$ runs are used to model facular regions of low and high activity respectively, and are hereafter referred to as weak-field and strong-field facular regions. We note that $\langle B_{\rm z} \rangle$ is the mean over the simulated cube and does not correspond to the magnetic field strengths measured for faculae in surface magnetograms.


The MURaM-derived facular region intensities are shown in Fig.~\ref{fig:hydroint} alongside the quiet star intensities. We provide \textit{Kepler}-band fit coefficients for the weak and strong-field facular regions in Table.~\ref{table:LD}, labelled $100\,{\rm G}$ and $500\,{\rm G}$ respectively. As the functional dependency between the quiet photosphere and facular region intensities ($I_{\rm phot}$, $I_{\rm fac}$) is difficult to distinguish in Fig.~\ref{fig:hydroint} (especially for later spectral types), we show facular intensity contrasts, $I_{\rm fac}/I_{\rm phot} - 1$ in Fig.~\ref{fig:LDcontrast} (top and middle panels). The weak-field facular regions (top panel) have positive contrasts for all spectral types and $\mu$, arising from facular brightening. They are brightest at the limbs of the stellar disc (low $\mu$) due to the limb-angle dependence of faculae \citep{Norris2017}. Strong-field facular regions (middle panel) have strong, positive contrasts at the limbs of the stellar disc for all spectral types. On Sun-like (G2 and K0) stars, facular region contrasts decrease to close to zero (cf. \citealp{Yeo2013}) at disc centre ($\mu=1.0$), whereas the M dwarf (M0 and M2) facular contrasts become negative, exhibiting spot-like behaviour near disc centre. This results from the presence of dark, pore-like structure in the simulated regions overwhelming the brightness contributions of the small-scale magnetic features forming the faculae.

In general, peak contrast levels for the strong-field regions are much higher than for the weak-field regions (roughly $5$ times higher for the G2 star). Both weak and strong-field facular contrasts tend to increase with stellar effective temperature (higher for earlier spectral types) at most limb distances. Solar facular contrasts \citep[derived from 1D empirical models,][]{Unruh1999} used in the Spectral And Total Irradiance REconstruction (SATIRE; see \citealp{Fligge2000} and \citealp{Krivova2003}) model are also plotted for comparison with the G2, strong-field MURaM results. The contrasts are of roughly the same magnitude for both models, though these 1D contrasts used in SATIRE are lower near the limb and slightly higher near disc centre.



Table \ref{table:muram} lists effective temperatures for the quiet star MURaM simulations $T_{\rm phot}$, as well as the effective temperature differences for the faculae $\Delta T_{\rm fac}\ (=T_{\rm fac} - T_{\rm phot})$ and spots (see Sect.~\ref{spotint}). The effective temperatures for the MURaM simulations (quiet stars and faculae) represent averages over several snapshots. They were derived from disc-integrated intensity spectra where Equation \ref{eq:limb_darkening} was used to extrapolate for $\mu < 0.2$. Temperatures calculated for the several simulation snapshots in each run differ due to oscillations but agree well overall, with standard deviations of the order of $6\,{\rm K}$ for the G2 and K0 runs, and $2\,{\rm K}$ for the M0 and M2. $T_{\rm phot}$ and $\Delta T_{\rm fac}$ are listed in Table \ref{table:muram}. $\Delta T_{\rm fac}$ is negative for the strong-field facular regions on M2 stars.

\subsection{Spot contrasts from 1D model atmospheres} \label{spotint}

\begin{table}
\caption{Stellar parameters used for all available spectral types in our simulations. $T_{\rm phot}$ are the equivalent effective temperatures of the quiet photosphere calculated from the MURaM-derived spectra. $\Delta T_{\rm fac}$ are the effective temperature differences from the photosphere calculated from MURaM $100\,{\rm G}$ and $500\,{\rm G}$ runs (for facular regions of weak and strong-field respectively). The spot temperature differences $\Delta T_{\rm spot}$ (based on \citealp{Panja2020}) and surface gravities $\log{g}$ are input parameters of the 1D radiative equilibrium model atmospheres used to calculate ATLAS9 spot spectra.}      
\label{table:muram} 
\centering             
\begin{tabular}{c c c c c c}  
\hline\hline                 
Type & $T_{\rm phot}$ [K] & \multicolumn{2}{c}{$\Delta T_{\rm fac}$ [K]} & $\Delta T_{\rm spot}$ [K] & $\log g$ \cr & & $100\,{\rm G}$ & $500\,{\rm G}$ & & \\  

\hline                        
   G2 &  5825 & 33 & 91 & 950 & 4.438\\
   K0 &  4904 & 21 & 31 & 561 & 4.609\\
   M0 &  3906 & 4 & 1 & 190 & 4.826\\
   M2 &  3675 & 2 & -13 & 80 & 4.826\\
\hline                                   
\end{tabular}
\end{table}


Sunspots are formed when a high concentration of magnetic field suppresses convective motion at the photosphere over a large horizontal area. If this area is sufficiently large, the reduction in convective energy transport is greater than the radiative flow through the side walls, resulting in a locally cool, dark region. On the Sun, spots are made up of two components: darker umbra at the centre surrounded by less-dark penumbra. Solar umbra and penumbra have effective temperatures $1000-1900\,{\rm K}$ and $250-400\,{\rm K}$ cooler than the quiet photosphere respectively \citep{Solanki2003}. Spot temperatures recorded for stars other than the Sun are subject to large uncertainties due to the limitations of (and discrepancies between) the indirect techniques required to observe them (see Appendix \ref{app:Tspot} for examples, and \citealp{Berdyugina2005} and \citealp{Strassmeier2009} for reviews). Spots have been simulated using the MURaM code for the Sun \citep{Rempel2009a, Rempel2009b} and other main sequence stars \citep{Panja2020}, providing umbral and penumbral effective temperatures for G2, K0 and M0 stars alongside other fundamental parameters. 

In our model, we do not distinguish between umbra and penumbra and instead assign an average effective temperature $T_{\rm spot}$ across whole spot areas, as in \citet{Chapman1987}, \citet{Unruh1999} and \citet{Herrero2016}. Solar irradiance variability modelling \citep{Wenzler2006} indicates that this approach does not significantly affect the resulting variability. We use umbral and penumbral temperatures based on the results of \citet{Panja2020} for G2, K0 and M0 stars and extrapolate temperatures for the M2 by fitting quadratic relations to the existing data. A 4:1 penumbral-to-umbral ratio is assumed to calculate the average effective spot temperature. For the spectral types considered in our models, the effective temperature differences between spots and the photosphere $\Delta T_{\rm spot}\ (=T_{\rm phot} - T_{\rm spot})$ are listed in Table \ref{table:muram}.



\begin{figure}
\centering
\includegraphics[width=\hsize]{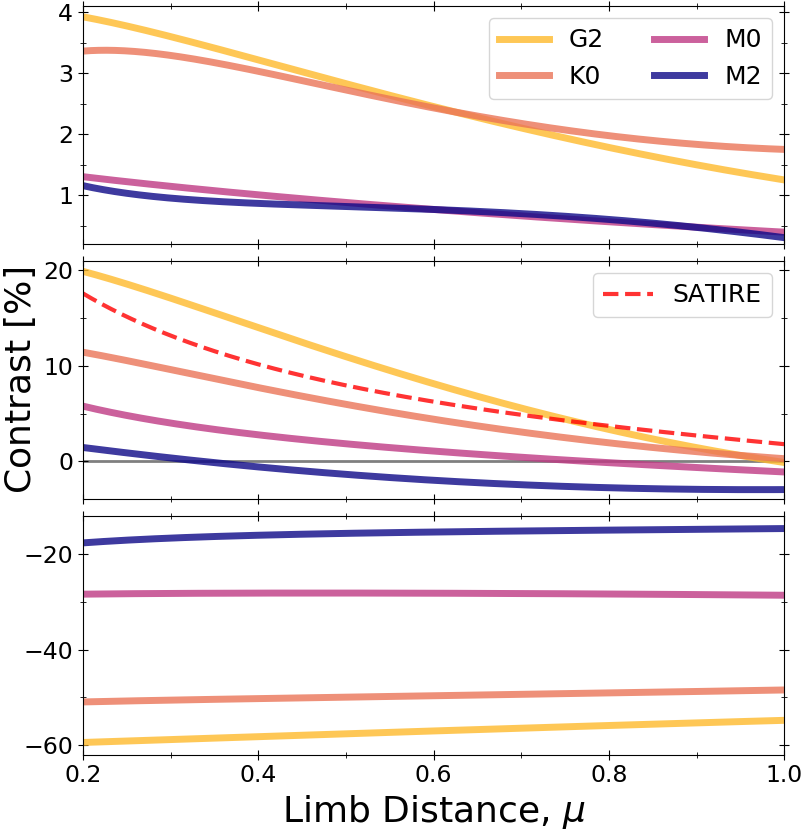}
\caption{\label{fig:LDcontrast}Intensity contrasts between photospheric features and the quiet photosphere against limb distance in the \textit{Kepler} band for G2, K0, M0 and M2 stars. Top and middle panels: MURaM-derived average facular region contrasts for runs with weak-field and strong-field facular regions, respectively. SATIRE model facular contrasts are also plotted \citep{Unruh1999}. Bottom panel: spot contrasts with temperature, $T_{\rm spot}$ (see Table \ref{table:muram}).}
\end{figure}


We follow \citet{Unruh2008} and use a 1D radiative equilibrium stellar atmosphere model with a cooler effective temperature than the quiet photosphere to represent spots. Mean MURaM quiet star spectra and 1D model spectra with equivalent effective temperature agree very well, although small differences are present for all spectral types. To minimise the effect of model differences on the resulting spot intensities, ATLAS9 spectra were calculated from 1D model stellar atmospheres at effective temperatures $T_{\rm phot}$ and $T_{\rm spot}$ for the quiet star and spots, respectively. The ratio of intensities was multiplied with the quiet star intensities from MURaM to provide spot intensities for our models. We show \textit{Kepler} band spot intensity contrasts, $I(T_{\rm spot})/I(T_{\rm phot}) - 1$ in Fig.~\ref{fig:LDcontrast} (bottom panel) for all spectral types. In the \textit{Kepler} band, spots always have negative contrasts several times larger than that of facular regions, regardless of limb distance. Spot contrasts become more negative with increasing stellar effective temperature (a direct consequence of the spot temperature differences chosen).

\section{Model setup} \label{distributions}

\begin{figure}
\includegraphics[width=\hsize]{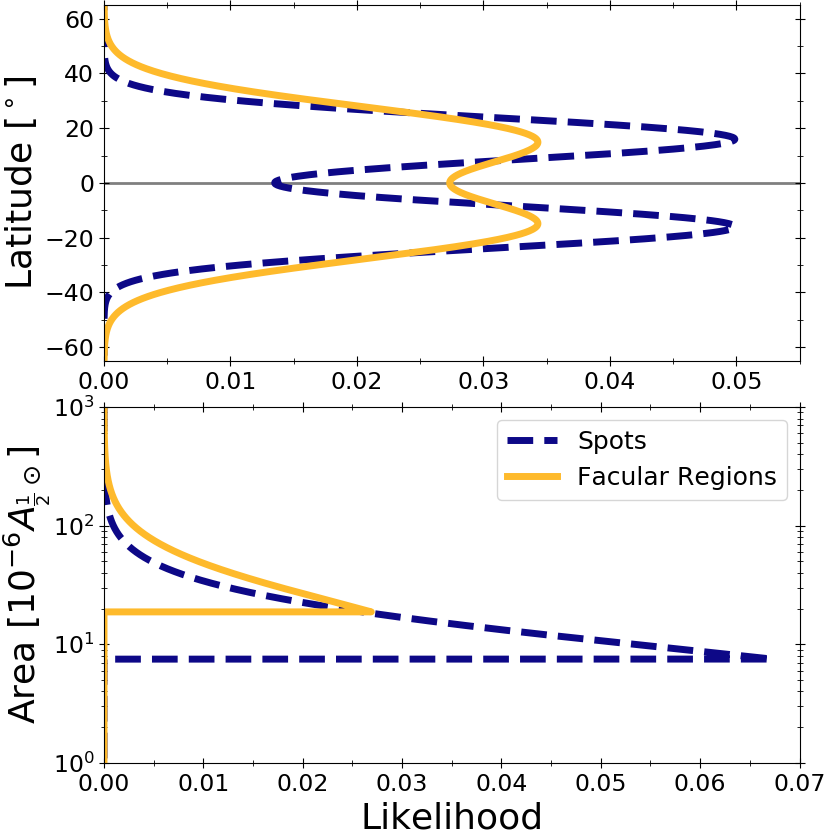}
\caption{\label{fig:Dists} Top panel: solar cycle average latitudinal distributions for spots (blue, dashed lines) and facular regions (yellow, solid lines) used in Sect.~\ref{MC}. Bottom panel: feature size distributions in units of $10^{-6}$ solar hemispheric area coverage, $A_{\frac{1}{2} \odot}$.}
\end{figure}

We performed a number of investigations using the lightcurve modelling approach described in Sect.~\ref{modelling}: obtaining typical variability levels assuming a Sun-like feature distribution and evaluating the impact of spot temperature, active region latitude and facular-to-spot coverage fraction on variability. Here, we describe how Sun-like feature distributions are generated and the assumptions made.


\subsection{Latitude and longitude distributions} \label{latlon}

We aim to model brightness variability on rotational timescales. Focusing on the Sun, spots and facular regions are observed to occur in latitudinal bands in both the northern and southern hemispheres. The central latitudes of these `active bands' are known to change throughout the solar activity cycle, beginning at higher latitudes at the start of each 11-year cycle and moving towards the equator as the cycle progresses  \citep{Hale1919}. To model this, we use typical solar cycle averaged latitudinal distributions. These distributions are based on \citet{Borgniet2015}, in which average spot and facular latitude distributions are presented for Solar Cycle 23 (May 1996 - October 2007), derived from USAF/NOAA sunspot group data and MDI/SOHO magnetograms \citep{Scherrer1995} respectively. Transit mapping observations of the K4V star HAT-P-11 show a very similar spot distribution to the Sun \citep{Morris2017}. This result indicates that the solar cycle averaged distributions in our simulations can also apply to some K-type stars.

We fitted the solar spot and facular region latitude data with equatorially symmetric Gaussian distributions (for spots, $\langle \phi \rangle = \pm 16^\circ$, $\sigma_{\phi} = 8^\circ$; for faculae, $\langle \phi \rangle = \pm 16^\circ$, $\sigma_{\phi} = 12^\circ$) shown in Fig.~\ref{fig:Dists} (top panel). Feature longitudes are randomised. As the number of spots (and the activity level) increases, this latitudinal distribution results in an active band between approximately $\pm 30^\circ$. If spots (or facular regions) completely filled this band, surface area coverage would be $50\%$ and disc coverage for a non-inclined star would be $61\%$. In the following we define the area coverages of spots $A_{\rm spot}$ and facular regions $A_{\rm fac}$ as percentages of the total stellar surface. 

\subsection{Size distributions} \label{SizeSect}

In \citet{Bogdan1988}, it was found that the size distribution of sunspots obeys a lognormal profile (cf. \citealp{Baumann2005}). For the size distribution of spots in our simulations, we use a lognormal distribution fitted by \citet{Solanki2004} to sunspot umbral areas recorded at Mt Wilson between $1921$ and $1982$ in units of parts-per-million (ppm) of a solar hemisphere's surface area ($10^{-6} A_{\frac{1}{2} \odot}$). The size distribution was multiplied by a factor of $5$ to represent total sunspot area (assuming a 4:1 penumbral-to-umbral area ratio) and the minimum spot size is set as $7.5 \times 10^{-6} A_{\frac{1}{2} \odot}$ \citep{Solanki2004}.

While individual faculae are much smaller than spots, solar faculae are grouped in active regions. Our models use average intensities from simulated `facular active regions' which are not completely filled with individual faculae. We obtain our facular region size distribution by multiplying the spot size distribution by an arbitrary factor of $2.5$. This rudimentarily mimicks how active regions observed on the solar surface emerge in groups, as well as increasing the computational efficiency of the simulation runs (by requiring fewer facular regions to reach the desired coverage level). The feature size distributions are shown in Fig.~\ref{fig:Dists} (bottom panel). 


\begin{figure*}
\centering
\includegraphics[width=\hsize]{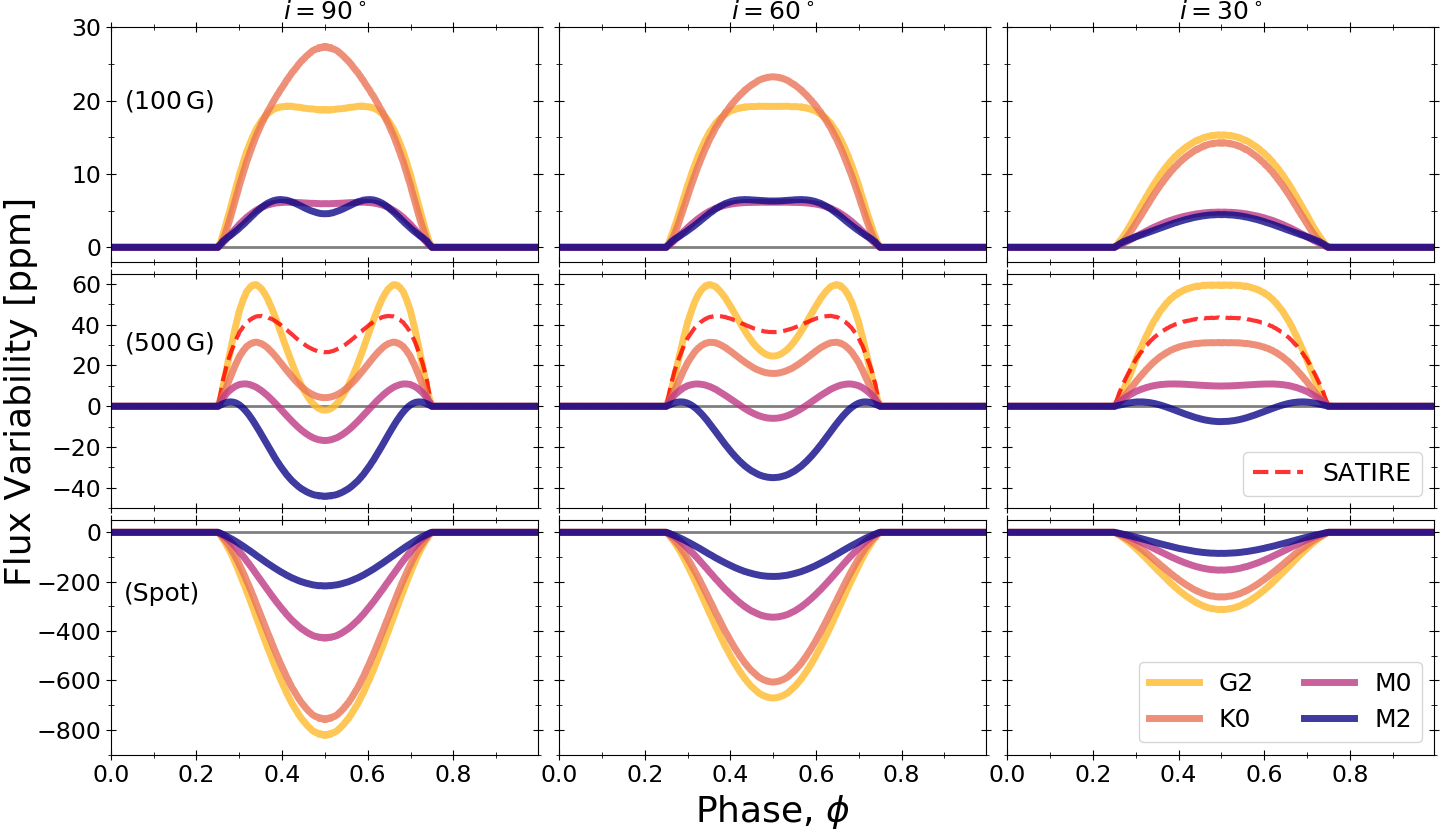}
\caption{\label{fig:LCsigs}\textit{Kepler}-band lightcurve signatures for a single equatorial feature with radius $2^\circ$, viewed equator-on at $i=90^\circ$ (left), $i=60^\circ$ (middle) and $i=30^\circ$ (right), for all available spectral types. Top and middle panels:  weak-field and strong-field facular regions respectively. SATIRE model facular lightcurve signatures are plotted alongside the MURaM strong-field signatures. Bottom panel: a spot with $T_{\rm spot}$ from Table \ref {table:muram}.}
\end{figure*}

\subsection{Facular-to-spot ratio and spatial association} \label{Qsec}

On the Sun, the area coverage of faculae, $A_{\rm fac, obs}$ is much greater than that of spots, $A_{\rm spot}$. The observed facular-to-spot coverage fraction, $Q_{\rm obs} = A_{\rm fac, obs}/A_{\rm spot}$ varies throughout the solar activity cycle: $Q_{\rm obs} \approx 10$ at activity maximum, and $Q_{\rm obs} > 40$ at activity minimum (\citealp{Chapman2001} and \citealp{Shapiro2014}). The value of $Q_{\rm obs}$ is heavily dependent on how solar faculae are identified.

As stated in Sect.~\ref{SizeSect}, we model facular active regions that are not completely filled with individual faculae. Thus, the facular-to-spot ratio $Q$ in our models is not equivalent to $Q_{\rm obs}$. Magnetic pixels in solar images from the Helioseismic and Magnetic Imager (HMI) onboard the Solar Dynamics Observatory \citep{Schou2012} are estimated to be $40\%$ filled with faculae (\citealp{Yeo2014}, \citealp{Chatzistergos2017} and \citealp{Chatzistergos2019}). We assume that MURaM boxes are equivalent to HMI magnetic pixels in facular filling, therefore using $Q = 10/0.4 = 25$ for solar activity maximum.


Following the spatial association of spots and faculae observed in solar images, we place half of the facular regions (by area) around spots. Every spot has a grouping of facular regions around it, with a coverage area roughly proportional to the spot area. The remaining facular regions are placed independently. A similar approach was implemented in \citet{Rackham2018, Rackham2019}.

\subsection{Monte-Carlo procedure} \label{mcp}

In our Monte-Carlo simulations, we sequentially added spots and facular regions to \texttt{actress} model stars, pseudo-randomly sampling feature latitudes, longitudes and sizes for $0 \leq A_{\rm spot} \leq 20\%$. Once added, the spots and facular regions do not evolve in time, remaining present throughout the time series of a simulated lightcurve (see \citealp{Nemec2020a, Nemec2020b} for solar lightcurve modelling accounting for the temporal evolution of magnetic features). Adding more features allows us to probe higher activity levels. An initial facular-to-spot coverage fraction $Q$ (which holds until $100\%$ coverage is reached) was used to update the facular region coverage $A_{\rm fac}$ after each spot is added and lightcurves were calculated at a series of regular intervals (finer gridding at low coverage levels\footnote{Sampling intervals of $\Delta A_{\rm spot} = 0.1\%$ up to $1.0\%$ spot coverage, $\Delta A_{\rm spot} = 0.2\%$ up to $3.0\%$, $\Delta A_{\rm spot} = 0.5\%$ up to $5.0\%$ and $\Delta A_{\rm spot} = 1.0\%$ up to $20.0\%$}) up to the maximum. The lightcurves were normalised by their means and the range variability $R_{\rm var}$ was calculated for each. We typically generated $N=320$ realisations for each parameter choice (all combinations of stellar spectral type and facular region field strength).

To allow the simulations to reach the desired coverage levels, the latitude constraint on facular regions is removed (enabling high latitude placement) once the total area coverage $A_{\rm tot} = (A_{\rm spot} + A_{\rm fac}) > 50\%$ (as stated in Sect.~\ref{latlon}, the low latitudes will be saturated with features at this coverage level). This is not entirely realistic as in practice, the whole active band latitude is expected to shift as activity level increases. High latitude facular region placement allows us to explore the fully-covered case and compare with other forward modelling studies (e.g. \citealp{Rackham2018, Rackham2019}). Once $A_{\rm tot} = 100\%$ is reached, $Q$ will decrease from its initial value as spots continue to be added. In addition, we begin gradually increasing the size distribution of facular regions once $A_{\rm tot} > 95\%$. This is because as the simulated stellar surface approaches complete coverage, the change in coverage with each new feature becomes increasingly small due to overlap with preexisting features. Increasing the feature sizes in this way ensures that total surface coverage is achieved within the alloted computation time. In this implementation, we do not account for superposition of features. The placement of a facular region will overwrite photosphere but not spots, and spot placement overwrites both photosphere and facular regions. See Appendix \ref{app:featdist} for further details on how the feature distributions change from low to high activity throughout a typical simulation run.


\section{Results} \label{results}

\subsection{Lightcurve signatures of single features} \label{LC}


To illustrate the lightcurve variability contributions of spots and facular regions, we set up a toy model consisting of a single equatorial feature (radius $r=2^\circ$) on an otherwise quiet stellar surface. This single feature was modelled as a spot and a facular region of weak or strong field, for all available spectral types in the \textit{Kepler} band. The resulting lightcurves (or feature signatures) are shown in Fig.~\ref{fig:LCsigs} for $i=90^\circ$, $60^\circ$ and $30^\circ$ and can be seen to reflect the intensity contrasts in Fig.~\ref{fig:LDcontrast}.

The weak-field facular region signatures (Fig.~\ref{fig:LCsigs}, top row) are observed to be bright for all spectral types. The signatures have flat-topped profiles for the G2 and M0 stars, double-peaked profiles for the M2 star at $i=90^\circ$ and $60^\circ$, and single-peaked profiles in all other cases. Decreasing the inclination generally reduces the signature sizes for the weak-field regions. This results from the equatorial feature being foreshortened towards the disc limbs (decreasing $\mu$) and therefore having a smaller projected area on the disc, reducing its brightness contribution. This effect dominates over the increased brightness contribution due to the facular regions having higher contrasts towards the limbs.

For the strong-field facular regions (Fig.~\ref{fig:LCsigs}, middle row), double-peaked signature profiles with central minima are observed for all spectral types when the star is viewed equator-on. For later spectral types, the peaks are less pronounced and spaced further apart and the central minimum has an increased depth. The Sun-like (G and K) facular signatures are brighter than the quiet photosphere everywhere except at disc centre. The M dwarf signatures exhibit brightening at the disc limbs, but are darker overall. As $i$ decreases, the central minima become shallower for all types while the peak heights are preserved. This is due to the regions being positioned further from disc centre, where contrasts are darkest. Lightcurve signatures are also calculated using SATIRE model intensities for comparison with our results. The SATIRE signatures have less pronounced peaks and central minima than the G2 signatures. 

For the spot regions, where we used the adopted relationship between the spot and photospheric temperature (Sect.~\ref{spotint}), lightcurve dips are larger for hotter stars (Fig.~\ref{fig:LCsigs}, bottom row). As for the weak-field facular regions, signature sizes decrease as $i$ decreases.


\subsection{Lightcurve variability simulations} \label{MC}

Monte-Carlo simulations were performed to investigate how lightcurve variability levels change with respect to stellar spectral type, feature area coverage, facular region field strength and inclination. We use a facular-to-spot area ratio of $Q=25$ here, corresponding to $Q_{\rm obs} = 10$, representative of solar activity maximum (see Sect.~\ref{Qsec}).
 
\begin{figure*}
\centering
\includegraphics[width=\hsize]{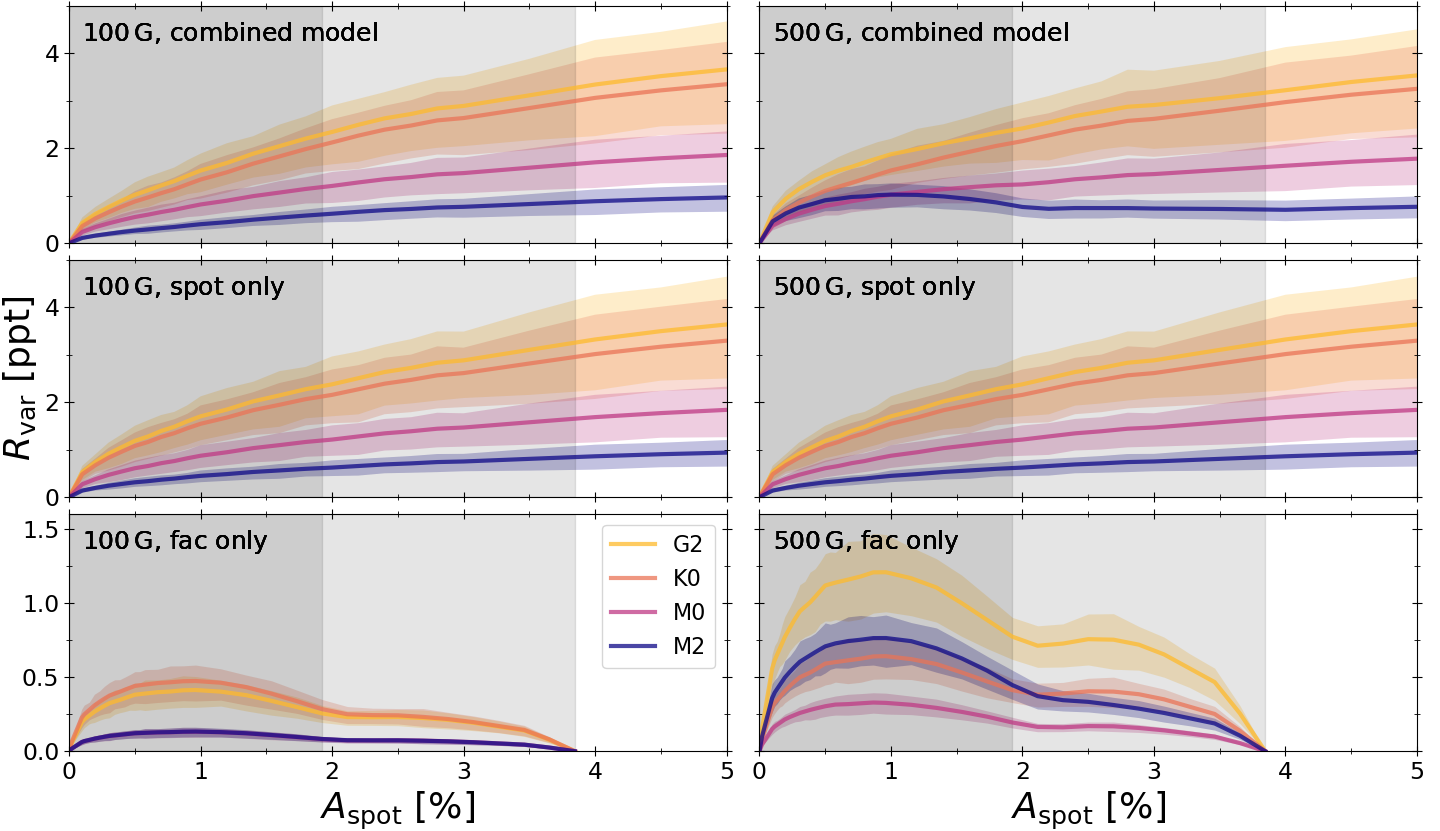}
\caption{\label{fig:Rvarsf}Range variability $R_{\rm var}$ against $A_{\rm spot}$ for the `combined', `spot only' and `facular only' models (top to bottom) for all spectral types at $i=90^\circ$ with weak-field (left) and strong-field facular regions (right). The two `spot only' panels are identical. Filled lines represent the means and shaded regions represent the interquartile ranges. Grey regions represent coverages $A_{\rm tot} \leq 50\%$ in which facular regions are latitudinally constrained, and $A_{\rm tot} \leq 100\%$ in which initial facular-to-spot ratio $(Q=25)$ holds.}
\end{figure*}

In Fig.~\ref{fig:Rvarsf}, $R_{\rm var}$ against spot coverage is shown for all available spectral types and facular region field strengths at stellar inclination $i=90^\circ$. The top, middle and bottom rows show the `combined', `spot only' and `facular only' models respectively, for spot filling factors up to $5\%$ of the total surface\footnote{The `facular only' results in Fig.\ref{fig:Rvarsf} share \textit{x}-axis units of $A_{\rm spot}$ with the `combined' and `spot only' results despite having no spots. This is done to allow easy comparison between the models. Here, facular coverage relates to spot coverage as $A_{\rm fac} = (Q+1) A_{\rm spot}$, where facular-to-spot ratio $Q=25$ in this case. The `$+1$' term results from facular regions previously overwritten by spots being revealed as the spots are removed, effectively replacing them (see Sect.~\ref{mcp}).} Grey regions on the plot represent total area coverages $A_{\rm tot} \leq 50\%$ in which facular regions are latitudinally constrained, and $A_{\rm tot} \leq 100\%$ in which initial $Q$ holds. Considering the `combined' model (top row), mean range variability as characterised by $R_{\rm var}$ increases steeply at first as $A_{\rm spot}$ increases, before tapering off. Variability levels have a clear spectral type dependence, with higher variability exhibited for earlier types. This is a direct consequence of the spot and facular contrasts, relative effective temperature differences $\Delta T_{\rm feature}/T_{\rm phot}$ (see Table~\ref{table:muram}) and bolometric intensity contrasts become smaller for cooler stars \citep{Panja2020}. We observed a large spread in variability around the mean in all cases (more so at higher coverage levels) and show the interquartile ranges to represent this\footnote{Interquartile ranges were plotted instead of 1-$\sigma$ intervals because the variability distributions are asymmetric and non-Gaussian.}. The spread is larger for earlier types (with higher variability). Increasing the facular field strength from weak (left column) to strong (right column) results in higher variability. The M2 star with strong-field facular regions exhibits a local maximum in variability around spot coverage $1\%$. 

The inclusion of facular regions in our simulations has a small but noticeable effect on range variability in the \textit{Kepler} band, shown here by comparing the `combined' model results to the `spot only' results (Fig.~\ref{fig:Rvarsf}, middle row). The inclusion of weak-field regions results in lower mean variability for all spectral types at coverages $A_{\rm spot} \lessapprox 2\%$. At higher coverage levels, variability is slightly higher than the `spot only' results for the Sun-like (G2 and K0) stars and roughly the same for the M-dwarfs (M0 and M2). For the strong-field case, the `combined' model results are higher than the `spot only' results for all spectral types at $A_{\rm spot} \lessapprox 3\%$, and lower than the `spot only' results at higher coverage levels. The local maximum in variability seen in the M2, strong-field case for the `combined' model at lower coverage is absent in the `spot only' results and is therefore a consequence of facular presence (M2 facular regions are dark, thus reinforcing rather than cancelling spot variability contributions).

The `facular only' results shown in the bottom row of Fig.~\ref{fig:Rvarsf} reveal that for all spectral types and facular field strengths, facular-induced variability increases with feature area coverage up to $A_{\rm spot} \approx 1\%$ (or $A_{\rm fac}\approx 25\%$), after which it begins to decrease again. This decrease results from the active latitude band (between approximately $\pm 30^\circ$) being over half-filled, therefore additional facular region placement will make the band more uniform. The secondary peaks result from relaxing the latitudinal constraint on facular region placement, for $A_{\rm fac} \approx 50\%$. At $A_{\rm spot} \approx 3.8\%$, the star is completely covered by facular regions and variability therefore decreases to zero. The facular variability contribution is greater with strong-field facular regions, though still lower than when spots are included in the model in most cases (the G2 `facular only' variability with strong-field regions is comparable to the `spot only' variability at $A_{\rm spot} \approx 1\%$).

\begin{figure}
\centering
\includegraphics[width=\hsize]{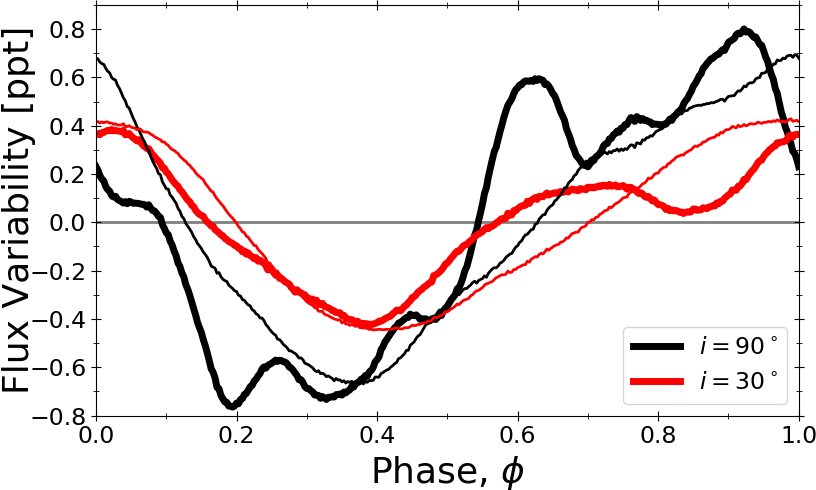}
\caption{\label{fig:lcso} \texttt{actress} example lightcurves for a G2 star with strong-field facular regions and feature coverage levels $A_{\rm spot} = 0.4\%$, $A_{\rm fac} = 10.0\%$ based on solar cycle maximum. Lightcurves calculated in the \textit{Kepler} band at inclinations $i=90^\circ$ (black line) and $30^\circ$ (red line), for the `combined' model (thick lines, from Fig.~\ref{fig:actsol}, middle panel) and the `spot only' model (thin lines). The lightcurves are mean-normalised.}
\end{figure}

While facular presence does not have a large impact on variability amplitudes (and therefore $R_{\rm var}$) in the \textit{Kepler} band, we found that inclusion of facular regions in our models dramatically influences the shape of lightcurves. Facular regions also affect the total brightness of the star (see Sect.~\ref{Qratio}). In Fig.~\ref{fig:lcso}, we show example lightcurves calculated with the `combined' model with strong-field facular regions, alongside lightcurves with spots only. The `combined' model lightcurves exhibit greater complexity, with more peaks and troughs than the `spot only' lightcurves. This is more so the case for the lightcurves calculated at stellar inclination $i=90^\circ$ than at $i=30^\circ$. Strong-field facular lightcurve signatures (see Fig.~\ref{fig:LCsigs}) have a high-frequency component at $i=90^\circ$ that is absent at $i=30^\circ$. Increased lightcurve complexity resulting from facular presence has the potential to complicate the determination of stellar rotation periods from photometry \citep{Shapiro2017, Shapiro2020}.

\subsection{Comparing simulations with observations} \label{obscomp}

\begin{figure*}
\centering
\includegraphics[width=\hsize]{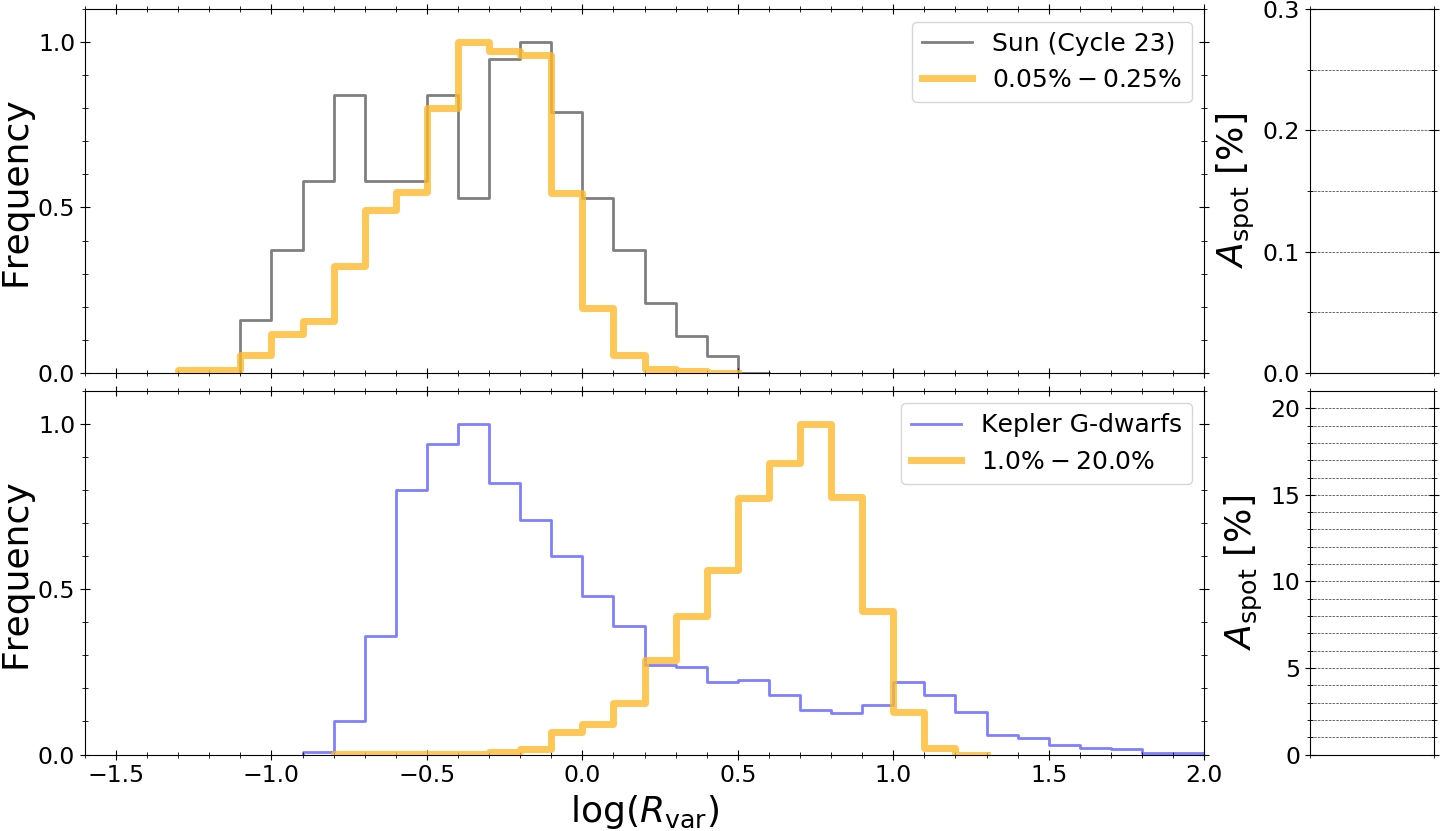}
\caption{\label{fig:hists}Histograms of $\log{R_{\rm var}}$ (with $R_{\rm var}$ in ppt) for a simulated G2 star at $i=90^\circ$ with weak-field facular regions, for all simulation runs with spot coverages $0.05 - 0.25\%$ (top panel) and $1.0-20.0\%$ (bottom panel). The variability over Solar Cycle 23 (calculated from a \textit{Kepler}-like solar lightcurve) and the \textit{Kepler} G-dwarf sample from \citet{Basri2013} are also plotted. The right-hand plots show the distribution of spot coverages of the simulation runs used (320 runs at each coverage level, represented by thin dashed lines) - density of lines gives the spot coverage distribution.
}
\end{figure*}

The spread in simulated variability around the mean is large at all coverage levels, as seen in Fig.~\ref{fig:Rvarsf}. We examine this further by performing a rudimentary comparison between simulated variability distributions and observations\footnote{In Sects.~$\ref{obscomp}-\ref{Qratio}$, a spot temperature of $4785\,{\rm K}$ is used for G2. This spot temperature is based on a linear fit to stellar observations (see Appendix \ref{app:Tspot}) and was used in this study before spot temperatures from \citet{Panja2020} were adopted. The difference between the two temperatures is small ($60\,{\rm K}$) and has a negligible effect on the investigation results.}. Fig.~\ref{fig:hists} shows histograms of $R_{\rm var}$ for the G2, weak-field facular region runs at $i=90^\circ$, for a range of spot coverages. The observed 30-day \textit{Kepler}-like variability over Solar Cycle 23 (the variability of a solar lightcurve created from composite VIRGO data, with a bolometric character comparable to \textit{Kepler} data) and of the \textit{Kepler} G-dwarf sample \citep[Fig.~4]{Basri2013} are plotted for comparison. On the right, the distribution of spot coverages for the simulation runs is shown. Each dashed line represents the centre of a spot coverage bin within which there are 320 simulation realisations. We note that the uniform distributions of spot coverages used have no physical basis (uniform coverage distributions are not seen throughout solar activity cycles, nor expected on other stars) and are used for illustrative purposes only.

In Fig.~\ref{fig:hists} (top panel), we show the simulated variability for spot coverages $0.05-0.25\%$, compared with the observed solar variability over Cycle 23. The high coverage cutoff in simulated spot coverage was chosen to reflect the observed maximal hemispheric area coverage from USAF/NOAA sunspot data for Cycle 23. The distributions appear slightly negatively skewed in log-space and therefore exhibit weak positive skewness in real space. This results from a higher probability of low-variability configurations in which features are near-uniformly distributed in longitude at all coverage levels. We also show the simulated variability for spot coverages $1.0-20.0\%$ in Fig.~\ref{fig:hists} (bottom panel), alongside the observed variability of the \textit{Kepler} G-dwarf sample. A low-coverage cutoff higher than for the Cycle 23 comparison was chosen because the Sun is thought to be relatively inactive compared to other solar-type stars (with known rotation periods close to solar, \citealp{Reinhold2020}). The high-coverage cutoff is also higher (the maximum spot coverage in our simulations).



\subsection{Effect of $T_{\rm spot}$ on lightcurve variability} \label{tspot}

The \textit{Kepler}-band range variability is very sensitive to changes in $T_{\rm spot}$ (at a fixed $T_{\rm phot}$). To investigate the effect of different spot effective temperatures, we calculated mean $R_{\rm var}$ of a G2 star for three different values. We chose $4041\,{\rm K}$ due to its use in \citet{Rackham2019} (representative of sunspot umbral temperatures) and $5100\,{\rm K}$ to represent average sunspot temperatures in agreement with SATIRE model fitting to SORCE/TIM data \citep{Ball2011, Ball2012}. $4785\,{\rm K}$ is the G2 spot temperature based on an unweighted linear fit to all stellar observations in Appendix \ref{app:Tspot}.

\begin{figure}
\centering
\includegraphics[width=\hsize]{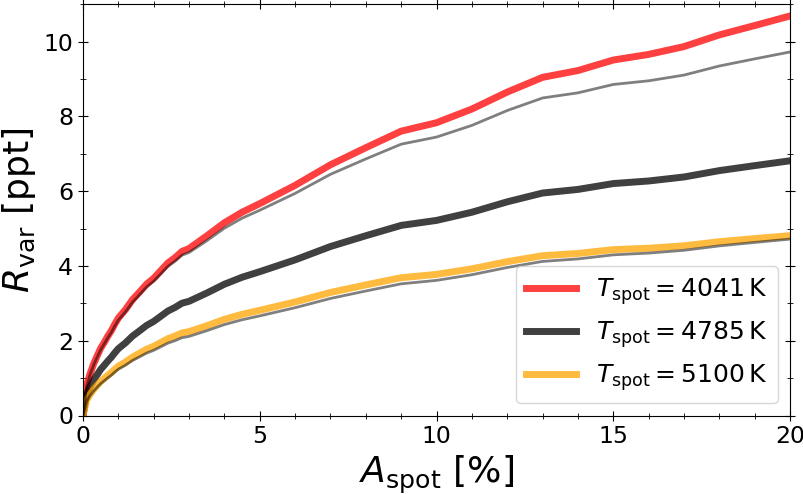}
\caption{\label{fig:RvarTspot}Mean $R_{\rm var}$ against $A_{\rm spot}$ for a G2 star with different spot temperatures, with weak-field facular regions at $i=90^\circ$. The thin grey lines are the $T_{\rm spot}=4785\,{\rm K}$ results multiplied by factors of $1.43$ and $0.69$.}
\end{figure}

Increasing $T_{\rm spot}$ (i.e. decreasing $\Delta T_{\rm spot}$) decreases the variability regardless of spectral type, facular field strength and stellar inclination angle $i$ (although only $i=90^\circ$ results are shown here, we found this behaviour to be unchanged at different $i$). Changes in $T_{\rm spot}$ result in an overall scaling of variability, with only small variations due to the variability contribution of the facular regions. This is illustrated by the thin grey lines in Fig.~\ref{fig:RvarTspot} which represent the $T_{\rm spot}=4785\,{\rm K}$ results scaled by factors of $1.43$ and $0.69$. These factors correspond to the \textit{Kepler}-band flux differences from the ATLAS9 spectra at $T_{\rm phot}$ and $T_{\rm spot}$. In the $T_{\rm spot}=5100\,{\rm K}$ model, scaling breaks down slightly at lower coverage levels ($A_{\rm spot} < 8.5\%$) as facular contributions become more important when $\Delta T_{\rm spot}$ is smaller. The difference between the $T_{\rm spot}=4041\,{\rm K}$ results and the scaled variability (greater at higher coverage levels) is a consequence of the maximum of the Planck function moving sufficiently to the red at low temperatures such that the linearity assumed for the scaling is no longer given.

\subsection{Effect of stellar inclination and active latitudes on lightcurve variability} \label{lats}

\begin{figure}
\centering
\includegraphics[width=\hsize]{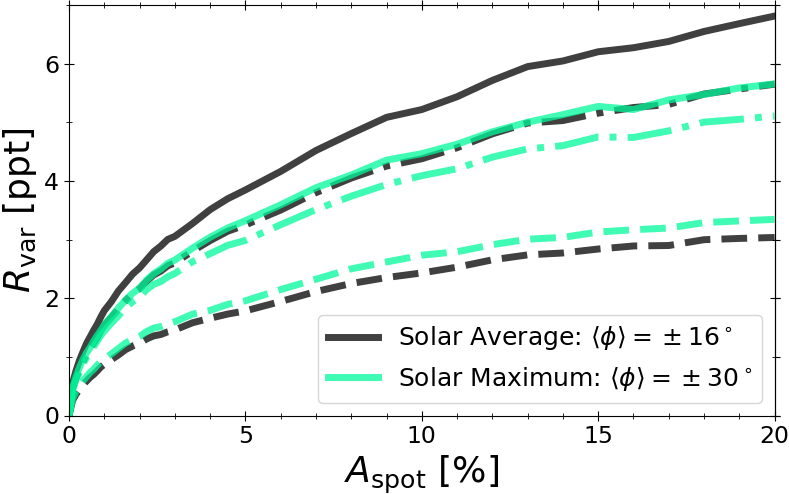}
\caption{\label{fig:Rvarlat}Mean $R_{\rm var}$ against $A_{\rm spot}$ for a G2 star with different active latitudes, at inclinations $i=90^\circ$ (filled lines), $60^\circ$ (dash-dotted lines) and $30^\circ$ (dashed lines) with weak-field facular regions.}
\end{figure}

In the initial simulation runs, lightcurves were simulated for stars viewed equator-on ($i=90^\circ$), with surface spots and facular regions distributed in active bands with central latitudes ($\phi = \pm 16^\circ$) based on the solar cycle average. Here, we investigate the effect of changing the stellar inclination angle and shifting the active bands on a G2 star further from the equator, with central latitudes ($\phi = \pm 30^\circ$) based on the maximum mean emergence latitude of the solar cycle. This typically occurs at the beginning of a cycle. The resulting plots are shown in Fig.~\ref{fig:Rvarlat}.

Comparing the mean $R_{\rm var}$ calculated for the `solar average' case at $i=90^\circ$, $60^\circ$ and $30^\circ$ reveals that variability decreases as inclination decreases. This results from the lightcurve signatures of spots and facular regions being smaller for lower $i$ (see Sect.~\ref{LC}). This behaviour is observed for all spectral types and facular region field strengths. Changing the active band latitudes also has a noticeable effect on the variability: increasing the central latitudes from $\pm 16^\circ$ to $\pm 30^\circ$ results in variability levels roughly $15\%$ lower on average at $i=90^\circ$, $8\%$ lower at $i=60^\circ$, and $10\%$ higher at $i=30^\circ$. This is because variability is highest when the active latitude band in which spots occur coincides with disc centre at a given stellar inclination. While spots are dominant in the \textit{Kepler} band, facular regions (which occur in broader latitude ranges than spots) also contribute. When the spot band coincides with disc centre, facular regions will be clustered nearer the `top' and `bottom' of the disc where contrasts are high, resulting in more high-variability configurations than otherwise. 

\subsection{Effect of $Q$ on lightcurve variability} \label{Qratio}

\begin{figure}
\centering
\includegraphics[width=\hsize]{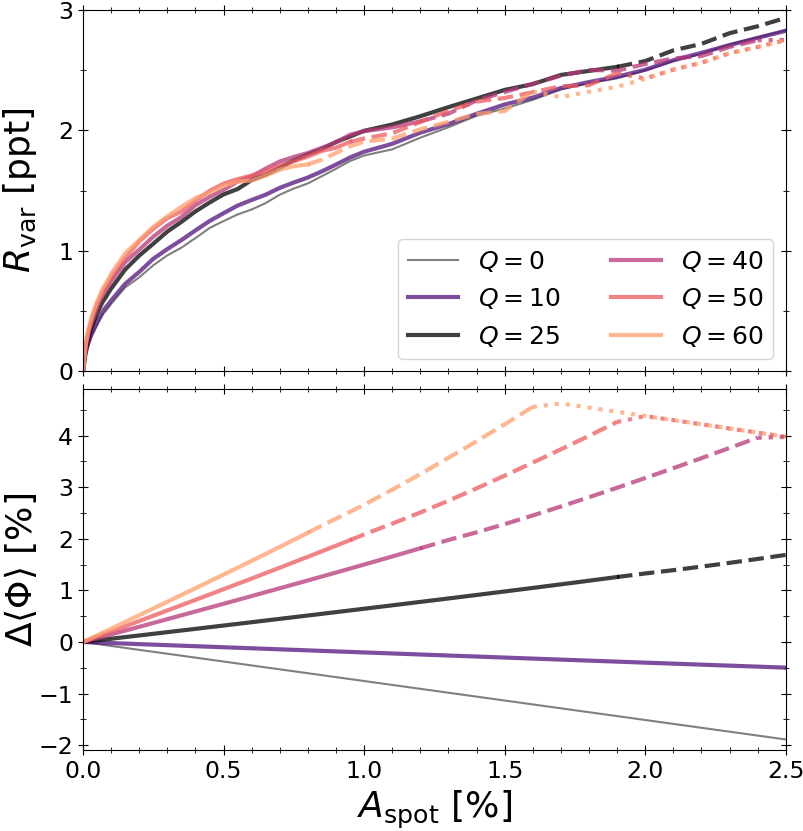}
\caption{\label{fig:RvarQ}Plots of $R_{\rm var}$ (top panel) and $\Delta \langle \Phi \rangle$ (bottom panel) against $A_{\rm spot}$ for a G2 star with facular-to-spot coverage fractions $Q=0$, $10$, $25$, $40$ and $60$, at $i=90^\circ$ with strong-field facular regions in the \textit{Kepler} band. $\Delta \langle \Phi \rangle$ is the relative change in mean brightness with respect to a featureless star. Solid lines represent total coverage levels at which facular regions are latitudinally constrained and initial $Q$ holds. Dashed and dotted lines represent total coverages $\geq 50\%$ at which the latitudinal constraint is relaxed, and dotted lines represent $100\%$ coverage at which $Q$ decreases due to spots overwriting facular regions. The solid, dashed and dotted lines are equivalent to the dark grey, light grey and unshaded regions in Fig.~\ref{fig:Rvarsf} respectively.}
\end{figure}

Here, we investigate the effect of varying the initial facular-to-spot area coverage fraction $Q$ on the variability. As stated in Sect.~\ref{Qsec}, the assumed relation between $Q$ and the observed solar facular-to-spot coverage fraction $Q_{\rm obs}$ is $Q \approx 2.5 \times Q_{\rm obs}$. The facular contribution to variability is largest for stars with relatively low activity levels and spot coverage fractions (\citealp{Lockwood2007}, \citealp{Hall2009} and \citealp{Radick2018}). We probe the low feature coverage range $A_{\rm spot} \leq 2.5\%$ with a much finer grid for a G2 star, viewed equator-on at $i=90^\circ$. The results for $Q=10$, $25$, $40$, $50$ and $60$ are plotted in Fig.~\ref{fig:RvarQ} for a G2 star with strong-field facular regions as both $R_{\rm var}$ (top panel) and as mean brightness normalised to the quiet star (bottom panel). 

Varying $Q$ has a relatively small effect on mean variability levels in the \textit{Kepler} band compared to other simulation parameters. For the small facular-to-spot ratios ($Q=10$), mean variability is slightly lower than the $Q=0$ (or `spot only') results. The high facular-to-spot ratio results ($Q=25$, $40$, $50$ and $60$) increases variability for $A_{\rm spot} \lessapprox 1.5\%$ and decreases variability at higher coverage levels. Typically, increasing $Q$ results in larger deviations from the `spot only' results. When the latitudinal constraint on facular regions is relaxed (illustrated by the change from solid to dashed lines), the variability slightly increases relative to what would otherwise be expected. This results from the previously unpopulated higher latitudes having inhomogeneities present. The local maximum in variability just before $Q$ begins to decrease (illustrated by the change from dashed to dotted lines, easiest seen in the $Q=60$ results at $A_{\rm spot} \approx 1.6\%$) is an artificial effect due to the increasingly large facular regions that are placed to ensure $100\%$ coverage is reached, resulting in some longitudes being much brighter than others before the photosphere is completely covered. The effects of varying $Q$ are similar for weak-field regions, though less pronounced.

As spots and facular regions are added to the simulated stellar surface, the mean brightness $\langle \Phi \rangle$ of the modelled stars changes. In Fig.~\ref{fig:RvarQ}, we show the change in mean brightness $\Delta \langle \Phi \rangle$ relative to the brightness of a featureless star for weak (top right) and strong-field facular regions (bottom right). In the presence of spots only ($Q=0$), the mean brightness decreases linearly with $A_{\rm spot}$ due to the increased presence of dark spots. The gradient (${\rm d} \langle \Phi \rangle/{\rm d} A_{\rm spot}$) becomes positive once both the facular-to-spot area ratio and facular field strength are sufficiently high for facular brightening to overcome spot darkening, resulting in an (initially linear) increase in mean brightness with spot coverage. A slight decrease in gradient is seen when the latitudinal constraint on facular regions is relaxed (dashed lines) (see Sect.~\ref{mcp}); this results from foreshortening of facular regions near the disc limbs overcoming the limb-brightening. The gradients sharply decrease when $Q$ begins to decrease as $100\%$ coverage is reached (dotted lines) and dark spots overwrite bright facular regions.

\subsection{\textit{TESS}-band variability} \label{tess}

\begin{figure}
\centering
\includegraphics[width=\hsize]{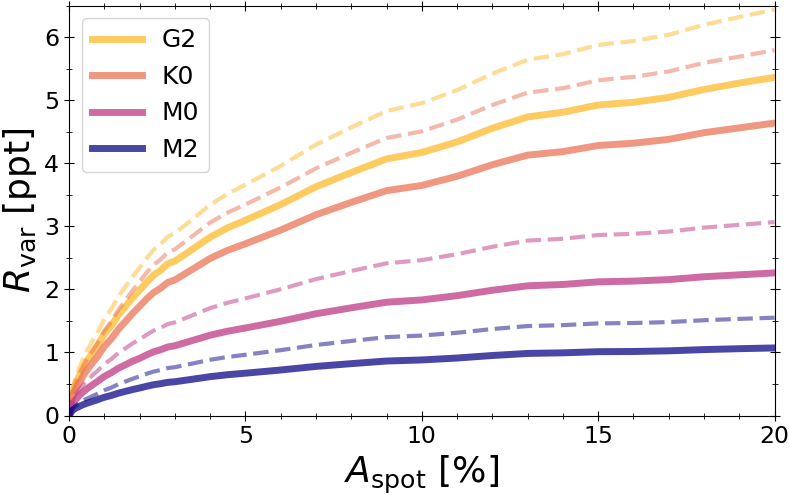}
\caption{\label{fig:TESS}Plot of mean range variability $R_{\rm var}$ against spot area coverage $A_{\rm spot}$ for all spectral types, in the \textit{TESS} band (filled lines) at stellar inclination $i=90^\circ$ with weak-field facular regions. The \textit{Kepler}-band results are plotted for comparison (thin, dashed lines).}
\end{figure}

The Transiting Exoplanet Survey Satellite (\textit{TESS}) launched in 2018, aiming to discover sub-Neptunes via transit photometry \citep{Ricker2015}. The \textit{TESS} detector bandpass spans roughly $600-1000\,{\rm nm}$, slightly redder and narrower than the \textit{Kepler} bandpass (roughly $400-900\,{\rm nm}$). In Fig.~\ref{fig:TESS}, we show the \textit{TESS}-band mean variability results calculated for spot coverage levels $0-20\%$, alongside the equivalent \textit{Kepler}-band results for all spectral types with weak-field facular regions (as in Sect.~\ref{MC}). Compared to the \textit{Kepler}-band results, \textit{TESS}-band variability is lower for all spectral types and coverage levels. This is a direct consequence of the spot and facular contrasts being smaller in redder wavelength intervals \citep{Nemec2020b}.

\section{Discussion} \label{disc}

Using \texttt{actress}, we calculated model \textit{Kepler}-band lightcurves for active, late-type stars of spectral type G2, K0, M0 and M2, featuring Sun-like surface distributions of spots and facular regions. We assume solid body rotation and model single period lightcurves only. The profiles of \texttt{actress} lightcurves result from the superposition of the signatures of all features on the model stellar photosphere. The intensity contrasts between the quiet star and surface features directly influence the size and shape of individual feature lightcurve signatures. Quiet star intensities and facular region contrasts are derived from MURaM 3D simulations of stellar surface magnetoconvection. Spot contrasts are calculated using ATLAS9 synthetic spectra with effective temperatures based on data from MURaM spot simulations \citep{Panja2020}. We calculate the proxy for variability, $R_{\rm var}$ as the outlier-trimmed amplitude of a lightcurve. As in \citet{Jackson2012, Jackson2013} and \citet{Rackham2018, Rackham2019}, we found that spot coverage level, spot temperature and feature distribution are the main drivers of \textit{Kepler}-band variability at a given stellar inclination. Variability levels are also affected by spectral type, facular region field strength and facular coverage level. 


Considering a star with only one type of surface feature (faculae or spots), as feature area coverage increases we found that mean $R_{\rm var}$ initially increases proportional to $\sqrt{A_{\rm feature}}$ before tapering off at higher coverage levels, eventually `turning over' and decreasing to zero. This is due to the fact that a half-filled active band will become more uniform as features continue to be added, reducing the variability. In \citet{Rackham2018, Rackham2019}, a comparable modelling approach is used: randomly distributing large spots with radius $2^\circ$ between latitudes $\pm 30^\circ$. The results are quantitatively similar to ours, finding an initial increase in variability with area coverage up to a turnover point, then decreasing again, reaching zero variability at $A_{\rm spot} \approx 33^\%$ when the active band is full (and therefore rotationally invariant). Using the same simulation parameters, we found the active band to be fully populated at $A_{\rm spot} \approx 50 \%$ instead. The key difference in our approaches is in modelling the geometry of the system. \texttt{actress} stellar discs are created using a single projected hemisphere of a HealPix 2-sphere, whereas \citet{Rackham2018, Rackham2019} apply a double cosine weighting kernel to half of a $180 \times 360$ pixel rectangular grid. This accounts for linear limb darkening and feature foreshortening in a rudimentary sense, but does not account for the surface area of a stellar hemisphere.


For individual lightcurves, the change in $R_{\rm var}$ when adding photospheric features is highly sensitive to the longitudinal placement. For example, a typical \textit{Kepler}-band lightcurve will have its largest peak corresponding to the longitude region containing the fewest spots. Placing a spot in this region will reduce the size of the lightcurve peak, therefore reducing $R_{\rm var}$. Placing a spot in the longitude region containing the highest density of spots will increase the size of the corresponding lightcurve trough, increasing $R_{\rm var}$. Depending on their placement and contrast, facular regions can enhance or diminish the variability. Mostly-bright facular regions (weak-field on all spectral types, strong-field on G and K stars) will increase the size of a lightcurve peak, or diminish the size of a lightcurve trough. The $50\%$ of facular regions placed near spots (see Sect.~\ref{Qsec}) will therefore generally lower the range variability. Mostly-dark strong-field features on M stars have a similar effect to spots, diminishing lightcurve peaks and accentuating troughs. This effect is seen in the M2 strong-field results in Fig.~\ref{fig:Rvarsf}, where mostly-dark facular regions amplify the variability at low coverage levels. We also find that including facular regions in our models dramatically changes the shape of lightcurves. Increased lightcurve complexity resulting from facular presence therefore has the potential to complicate the determination of stellar rotation periods from photometry \citep{Shapiro2017, Shapiro2020}.

While the sensitivity of individual lightcurve variability to the longitudinal distribution of features contributes significantly to the large spread in $R_{\rm var}$ around the mean, the mean itself is mainly dependent on the location of the active latitudes (see Fig.~\ref{fig:Rvarlat}). Using a Sun-like latitude distribution (active bands centred on $\phi = \pm 16^\circ$), $R_{\rm var}$ is highest when the star is viewed equator-on ($i=90^\circ$). This effect was also seen by \citet{Meunier2019b} who analysed synthetic activity time series for F6-K4 stars. Shifting the active band latitudes to $\phi = \pm 30^\circ$ causes $R_{\rm var}$ to decrease when $i=90^\circ$, and increase when $i=60^\circ$. This is caused by the active latitudes being positioned closer to disc centre when the star is inclined. While high latitude features like this are not seen on the Sun, Doppler images of faster rotating stars \citep{Strassmeier2009} reveal high latitude \citep{Wolter2005, Wolter2008} and even polar (\citealp{Barnes2001b} and \citealp{Barnes2005}) feature emergence. This can arise from the polewards deflection of the buoyant magnetic flux tubes (that form spots and facular regions at the photosphere) as they rise through the outer convective envelope of the star \citep{Schussler1996}. Coupling our model with  simulations of the rise of magnetic flux through the stellar convection zone \citep[e.g.][]{Isik2018} would allow us to implement realistic feature distributions for stars more active than the Sun, as well as providing spot and facular region filling factors.

Varying the facular-to-spot coverage fraction in the range $0 \leq Q \leq 60$ was found to have a relatively small, but significant effect on mean $R_{\rm var}$, compared to the other parameters probed in our simulations (see Fig.~\ref{fig:RvarQ}, top panel). We found that, for $Q=25$, $40$, $50$ and $60$ and in the coverage range probed, the inclusion of facular regions in our models results in increased $R_{\rm var}$ at coverages below $A_{\rm spot} \approx 1.5\%$, and lower variability at higher coverages. This can be attributed to large, bright regions increasing lightcurve amplitude up until the photosphere is over half-covered. For the $Q=10$ runs, the presence of facular regions reduces variability levels at all coverage levels. This results from the spatial association between spots and facular regions in our simulations ensuring a concentration of bright regions around the dark spots, reducing lightcurve amplitudes. The increased coverage of unassociated facular regions at high $Q$ will result in more high-variability configurations. However, as $R_{\rm var}$ is calculated from mean-normalised lightcurves, facular brightening leads to a slight reduction in lightcurve amplitudes overall, lessening the aforementioned effect.

The high sensitivity of the variability proxy $R_{\rm var}$ to spot temperature in our simulations (see Fig.~\ref{fig:RvarTspot}) highlights the importance of using an accurate $T_{\rm spot}$ determination for each spectral type. Underestimating spot temperatures (and therefore overestimating their variability contribution) might wrongly suggest that faculae can be neglected. The $T_{\rm spot}$ to $T_{\rm phot}$ relationship implemented in this paper is based on data from MURaM simulations \citep{Panja2020}. In Appendix \ref{app:Tspot}, we show results based on an unweighted linear fit to stellar observations, the majority of which were determined using TiO-band modelling and lightcurve analysis methods. Temperature determinations with these methods are highly uncertain and often inconsistent with each other. This inconsistency results from the different methods often reflecting different physical temperatures; TiO-band intensity is weakly temperature dependent at $T \geq 3800\,{\rm K}$ (\citealp{Neff1995} and \citealp{ONeal1996}) and is therefore sensitive to spots on late-K and M dwarfs, but only umbral temperatures on hotter stars. Lightcurves are sensitive to umbra and penumbra regardless of stellar effective temperature. In addition, there exists a degeneracy between contrast and filling factor in lightcurves that cannot be broken when single bands (e.g. \textit{Kepler}) are used. TiO temperature determinations for spots (on Sun-like stars especially) appear to be consistently lower than those obtained through lightcurve analysis. More accurate spectral determination of starspot temperatures is becoming possible by modelling spot occultations during planetary transits in spectroscopic observations \citep[see][]{Espinoza2019}. Effective spot temperatures determined using MURaM simulations show good agreement with the solar case and the relationship between spot and photospheric temperature has a similar gradient to the \citet{Rackham2019} fit to the dwarf star data in \citet{Berdyugina2005}.

The `facular only' case was investigated (see Fig.~\ref{fig:Rvarsf}, bottom row), showing that the removal of the high-contrast spots results in a dramatic reduction in range variability at all coverage levels. On the Sun, high facular-to-spot coverage fractions $Q_{\rm obs}>40$ are observed during low activity periods of the solar cycle. In some periods, the solar surface has been observed to be devoid of spots (while still populated with faculae). During activity maxima, $Q_{\rm obs}$ decreases dramatically ($Q_{\rm obs} \approx 10$) due to the increased presence of spots.

We ascertained from varying $Q$ that the range variability $R_{\rm var}$ is a poor probe of facular presence. However, the mean brightness (and lightcurve shape) is significantly affected by faculae. Understanding how facular presence affects mean brightness is an important consideration when determining planet sizes via transit photometry. As would be expected, a linear change in mean brightness with spot and facular region coverage is seen when the facular-to-spot ratio is constant (see Fig.~\ref{fig:RvarQ}, right). For G and K stars with typically bright facular regions (regardless of facular field strength), the higher the facular-to-spot ratio, the greater the differential increase in brightness. The inverse is true for M stars with strong-field facular regions that appear dark relative to the photosphere at most limb distances.

Our simulation results for the G2 star reveal that mean brightness increases with coverage level for $Q=25$ with strong-field facular regions, and $Q=40$, $50$ and $60$ regardless of facular field strength. The Sun exhibits a strong activity-brightness correlation; at activity maximum, TSI (Total Solar Irradiance) is higher due to the facular brightening overcoming spot darkening. Comparisons between the chromospheric activity and photometric variability of solar analogs \citep{Hall2009} have revealed that they typically exhibit at most a weak activity-brightness correlation. This could result from their higher average activity levels relative to the Sun, therefore having higher spot coverages and lower facular-to-spot ratios. 

Comparing \texttt{actress}-simulated variability for a solar type star (viewed equator-on with weak-field facular regions, spot coverages uniformly distributed between $0.05-0.25\%$ and $Q=25$) and the 30-day variability measured over Cycle 23 \citep{Basri2013} reveals that the range of variability and typical variability levels are of the same order (see Fig.~\ref{fig:hists}, top). Throughout a typical solar cycle, there are more periods of relatively low sunspot area coverage than otherwise. This explains why the simulated distribution (with uniformly distributed coverages) has fewer low-variability occurrences than the observed distribution. The simulated distribution also has fewer high-variability occurrences: observed solar surface magnetic fields (and the emergent features) are rarely distributed uniformly in longitude, instead concentrating in `clumps' and therefore amplifying the variability. This effect is explored through lightcurve modelling in \citet{Isik2020} and could be approximated by implementing `active longitude' bands in the simulations.


We also compared G2 simulations (spot coverage range $1.0-20.0\%$, same simulation parameters as in the solar comparison) with the observed variability of the \textit{Kepler} G-dwarf sample \citep{Basri2013} (see Fig.~\ref{fig:hists}, bottom). The observations have a much stronger positive skew than the simulated data, suggesting that there is a much higher proportion of relatively inactive stars in the sample than is reflected by the uniform spot coverage distribution used. In addition, the range of stellar inclinations in the G-dwarf sample will result in lower variability overall, as variability is maximised (in the \textit{Kepler} band) when the star is viewed close to equator-on. The high-variability tail in the observed data cannot be explained by higher spot coverages alone; the results in Sect.~\ref{MC} indicate that the increase in range variability with spot coverage lessens at higher coverage levels. Like the high-variability occurrences in the solar data, this could result from some longitudinal regions having high concentrations of magnetic activity.

A more complete comparison between simulated and observed data will be conducted in future: applying our understanding of the solar activity cycle (how feature coverage levels and emergence latitudes vary throughout), updating simulation parameters such as spot temperature and implementing active longitude bands. We also plan to investigate the wavelength dependence of activity-induced variability: solar facular contrasts are greatly enhanced in UV and are comparable to spot contrasts \citep[see][and references therein]{Yeo2013}, whereas contrasts of both the facular regions and spots are greatly diminished in the IR.

In this study, we consider stellar variability on rotational timescales only (days to months), where spots and faculae dominate. Another source of variability on these timescales is the evolution of magnetic features, which we do not consider at present (see \citealp{Nemec2020a, Nemec2020b} for solar variability modelling with feature evolution). Variability on shorter timescales (minutes to hours) results mainly from granulation at the photosphere and pressure modes. Instrumental noise also contributes to observed variability on shorter timescales \citep{Shapiro2017}. Long-term variability (years to decades) is a consequence of the stellar activity cycle, throughout which typical levels of spot and facular surface coverage fluctuate \citep{Solanki2013}.


\section{Conclusions} \label{conc}

In this paper, we present \texttt{actress}, a geometrically accurate model for the stellar variability of late-type stars due to faculae and spots on rotational timescales, without spot evolution. The advantage of our modelling approach over previous works lies in the treatment of facular active regions (on the stellar photosphere). Intensity spectra calculated from MURaM 3D magnetoconvection simulations for G2, K0, M0 and M2 stars are used to derive limb-dependent intensities for the quiet photosphere and facular regions in a chosen spectral band. We provide the limb-dependent intensity coefficients used in this work for the \textit{Kepler} band (\textit{TESS}-band coefficients are also provided, see Appendix \ref{app:tessfit}).



We calculate single-phase \textit{Kepler}-band lightcurves in Monte-Carlo simulations, investigating the sensitivity of the lightcurve variability proxy $R_{\rm var}$ to changes in simulation parameters (spectral type, stellar inclination, feature distributions, spot temperature, facular field strength, spot area coverage and facular-to-spot coverage fraction). Our results show good qualitative agreement with prior studies. We attribute the quantitative differences to our accurate geometric setup and prescribed limb-dependent intensities. We found that, for a given spectral type and stellar inclination, spot temperature and spot area coverage have the largest effect on $R_{\rm var}$ of all simulation parameters. This confirms the spot-dominated nature of variability in the \textit{Kepler} band.

The inclusion of facular regions in our models (and varying facular simulation parameters) has a relatively small influence on $R_{\rm var}$. This does not indicate that faculae are unimportant, but rather that $R_{\rm var}$ is not a good indicator of facular coverage. We found that faculae have a strong influence on mean brightness levels and on the lightcurve shape, potentially inhibiting planet size determination via transit photometry and stellar periodicity measurements respectively.


\section*{Acknowledgements}
We thank Emre I{\c{s}}ık, Theodosios Chatzistergos, Mayukh Panja, James Owen and Subhanjoy Mohanty for helpful advice and fruitful discussion on the topic of solar and stellar variability. We also thank the anonymous referee for their useful
comments which helped to improve the paper. This work was supported through studentship funding of the UK Science and Technology Facilities Council (STFC). YCU acknowledges funding through STFC consolidated grants ST/S000372/1 and ST/N000838/1. This work has been partially supported by the BK21 plus program through the National Research Foundation (NRF) funded by the Ministry of Education of Korea. VW and AIS acknowledge funding from the European Research Council under the European Union Horizon 2020 research and innovation programme (grant agreement No. 715947). The authors thank the International Space Science Institute, Bern, for their support of Science Team 446 and the resulting helpful discussions.

\section*{Data availability}
The data underlying this article are available upon reasonable request.


\bibliographystyle{mnras} 
\bibliography{ref} 

\begin{thebibliography}{}
\makeatletter
\relax
\def\mn@urlcharsother{\let\do\@makeother \do\$\do\&\do\#\do\^\do\_\do\%\do\~}
\def\mn@doi{\begingroup\mn@urlcharsother \@ifnextchar [ {\mn@doi@}
  {\mn@doi@[]}}
\def\mn@doi@[#1]#2{\def\@tempa{#1}\ifx\@tempa\@empty \href
  {http://dx.doi.org/#2} {doi:#2}\else \href {http://dx.doi.org/#2} {#1}\fi
  \endgroup}
\def\mn@eprint#1#2{\mn@eprint@#1:#2::\@nil}
\def\mn@eprint@arXiv#1{\href {http://arxiv.org/abs/#1} {{\tt arXiv:#1}}}
\def\mn@eprint@dblp#1{\href {http://dblp.uni-trier.de/rec/bibtex/#1.xml}
  {dblp:#1}}
\def\mn@eprint@#1:#2:#3:#4\@nil{\def\@tempa {#1}\def\@tempb {#2}\def\@tempc
  {#3}\ifx \@tempc \@empty \let \@tempc \@tempb \let \@tempb \@tempa \fi \ifx
  \@tempb \@empty \def\@tempb {arXiv}\fi \@ifundefined
  {mn@eprint@\@tempb}{\@tempb:\@tempc}{\expandafter \expandafter \csname
  mn@eprint@\@tempb\endcsname \expandafter{\@tempc}}}

\bibitem[\protect\citeauthoryear{Afram, Unruh, Solanki, Sch{\"{u}}ssler, Lagg
  \& V{\"{o}}gler}{Afram et~al.}{2011}]{Afram2011}
Afram N.,  Unruh Y.~C.,  Solanki S.~K.,  Sch{\"{u}}ssler M.,  Lagg A.,
  V{\"{o}}gler A.,  2011, \mn@doi [A{\&}A] {10.1051/0004-6361/201015582}, 526,
  A120

\bibitem[\protect\citeauthoryear{Andersen \& Korhonen}{Andersen \&
  Korhonen}{2015}]{Andersen2015}
Andersen J.~M.,  Korhonen H.,  2015, \mn@doi [MNRAS] {10.1093/mnras/stu2731},
  448, 3053

\bibitem[\protect\citeauthoryear{Ball, Unruh, Krivova, Solanki  \& Harder}{Ball
  et~al.}{2011}]{Ball2011}
Ball W.~T.,  Unruh Y.~C.,  Krivova N.~A.,  Solanki S.,   Harder J.~W.,  2011,
  A{\&}A, 530, A71

\bibitem[\protect\citeauthoryear{Ball, Unruh, Krivova, Solanki, Wenzler,
  Mortlock  \& Jaffe}{Ball et~al.}{2012}]{Ball2012}
Ball W.~T.,  Unruh Y.~C.,  Krivova N.~A.,  Solanki S.,  Wenzler T.,  Mortlock
  D.~J.,   Jaffe A.~H.,  2012, \mn@doi [A{\&}A] {10.1051/0004-6361/201118702},
  541, A27

\bibitem[\protect\citeauthoryear{Barnes}{Barnes}{2005}]{Barnes2005}
Barnes J.~R.,  2005, \mn@doi [MNRAS] {10.1111/j.1365-2966.2005.09544.x}, 364,
  137

\bibitem[\protect\citeauthoryear{Barnes, Collier~Cameron, James  \&
  Steeghs}{Barnes et~al.}{2001}]{Barnes2001b}
Barnes J.~R.,  Collier~Cameron A.,  James D.~J.,   Steeghs D.,  2001, \mn@doi
  [MNRAS] {10.1046/j.1365-8711.2001.04648.x}, 326, 1057

\bibitem[\protect\citeauthoryear{Basri \& Nguyen}{Basri \&
  Nguyen}{2018}]{Basri2018}
Basri G.,  Nguyen H.~T.,  2018, \mn@doi [ApJ] {10.3847/1538-4357/aad3b6}, 863,
  190

\bibitem[\protect\citeauthoryear{Basri et~al.,}{Basri et~al.}{2011}]{Basri2011}
Basri G.,  et~al., 2011, \mn@doi [AJ] {10.1088/0004-6256/141/1/20}, 141, 20

\bibitem[\protect\citeauthoryear{Basri, Walkowicz  \& Reiners}{Basri
  et~al.}{2013}]{Basri2013}
Basri G.,  Walkowicz L.~M.,   Reiners A.,  2013, \mn@doi [ApJ]
  {10.1088/0004-637X/769/1/37}, 769, A37

\bibitem[\protect\citeauthoryear{Baumann \& Solanki}{Baumann \&
  Solanki}{2005}]{Baumann2005}
Baumann I.,  Solanki S.~K.,  2005, \mn@doi [A{\&}A]
  {10.1051/0004-6361:20053415}, 443, 1061

\bibitem[\protect\citeauthoryear{Beeck, Cameron, Reiners  \&
  Sch{\"{u}}ssler}{Beeck et~al.}{2013}]{Beeck2013a}
Beeck B.,  Cameron R.~H.,  Reiners A.,   Sch{\"{u}}ssler M.,  2013, A{\&}A,
  558, A48

\bibitem[\protect\citeauthoryear{Beeck, Sch{\"{u}}ssler, Cameron  \&
  Reiners}{Beeck et~al.}{2015}]{Beeck2015b}
Beeck B.,  Sch{\"{u}}ssler M.,  Cameron R.~H.,   Reiners A.,  2015, A{\&}A,
  581, A42

\bibitem[\protect\citeauthoryear{Berdyugina}{Berdyugina}{2004}]{Berdyugina2004}
Berdyugina S.~V.,  2004, \mn@doi [Solar Phys.] {10.1007/s11207-005-6503-3},
  224, 123

\bibitem[\protect\citeauthoryear{Berdyugina}{Berdyugina}{2005}]{Berdyugina2005}
Berdyugina S.~V.,  2005, Living Rev. Solar Phys., 2, 62

\bibitem[\protect\citeauthoryear{Bogdan, Gilman, Lerche  \& Howard}{Bogdan
  et~al.}{1988}]{Bogdan1988}
Bogdan T.~J.,  Gilman P.~A.,  Lerche I.,   Howard R.,  1988, \mn@doi [ApJ]
  {10.1086/166206}, 327, 451

\bibitem[\protect\citeauthoryear{Borgniet, Meunier  \& Lagrange}{Borgniet
  et~al.}{2015}]{Borgniet2015}
Borgniet S.,  Meunier N.,   Lagrange A.~M.,  2015, \mn@doi [A{\&}A]
  {10.1051/0004-6361/201425007}, 581, A133

\bibitem[\protect\citeauthoryear{Borucki et~al.,}{Borucki
  et~al.}{2010}]{Borucki2010}
Borucki W.~J.,  et~al., 2010, \mn@doi [Science] {10.1126/science.1185402}, 327,
  977

\bibitem[\protect\citeauthoryear{Carlsson, Stein, Nordlund  \&
  Scharmer}{Carlsson et~al.}{2004}]{Carlsson2004}
Carlsson M.,  Stein R.~F.,  Nordlund r.,   Scharmer G.~B.,  2004, \mn@doi [ApJ]
  {10.1086/423305}, 610, L137

\bibitem[\protect\citeauthoryear{Castelli \& Kurucz}{Castelli \&
  Kurucz}{2003}]{Castelli2003}
Castelli F.,  Kurucz R.~L.,  2003, in Piskunov N.,  Weiss W.~W.,   Gray D.~F.,
  eds,  IAU Symposium Vol. 210, Modelling of Stellar Atmospheres. p.~A20

\bibitem[\protect\citeauthoryear{Cauley, Redfield  \& Jensen}{Cauley
  et~al.}{2017}]{Cauley2017}
Cauley P.~W.,  Redfield S.,   Jensen A.~G.,  2017, \mn@doi [AJ]
  {10.3847/1538-3881/aa6a15}, 153, 217

\bibitem[\protect\citeauthoryear{Chapman}{Chapman}{1987}]{Chapman1987}
Chapman G.~A.,  1987, \mn@doi [ARA{\&}A] {10.1146/annurev.aa.25.090187.003221},
  25, 633

\bibitem[\protect\citeauthoryear{Chapman, Cookson, Dobias  \& Walton}{Chapman
  et~al.}{2001}]{Chapman2001}
Chapman G.~A.,  Cookson A.~M.,  Dobias J.~J.,   Walton S.~R.,  2001, \mn@doi
  [ApJ] {10.1086/321466}, 555, 462

\bibitem[\protect\citeauthoryear{Chatzistergos}{Chatzistergos}{2017}]{Chatzistergos2017}
Chatzistergos T.,  2017, PhD thesis, Max Planck Institute for Solar System
  Research, G{\"{o}}ttingen, Germany

\bibitem[\protect\citeauthoryear{Chatzistergos, Ermolli, Krivova  \&
  Solanki}{Chatzistergos et~al.}{2019}]{Chatzistergos2019}
Chatzistergos T.,  Ermolli I.,  Krivova N.~A.,   Solanki S.~K.,  2019, \mn@doi
  [A{\&}A] {10.1051/0004-6361/201834402}, 625, A69

\bibitem[\protect\citeauthoryear{Czesla, Huber, Wolter, Schr{\"{o}}ter  \&
  Schmitt}{Czesla et~al.}{2009}]{Czesla2009}
Czesla S.,  Huber K.~F.,  Wolter U.,  Schr{\"{o}}ter S.,   Schmitt J. H. M.~M.,
   2009, \mn@doi [A{\&}A] {10.1051/0004-6361/200912454}, 505, 1277

\bibitem[\protect\citeauthoryear{Danilovic, Sch{\"{u}}ssler  \&
  Solanki}{Danilovic et~al.}{2010}]{Danilovic2010}
Danilovic S.,  Sch{\"{u}}ssler M.,   Solanki S.~K.,  2010, \mn@doi [A{\&}A]
  {10.1051/0004-6361/200913379}, 513, A1

\bibitem[\protect\citeauthoryear{Desort, Lagrange, Galland, Udry  \&
  Mayor}{Desort et~al.}{2007}]{Desort2007}
Desort M.,  Lagrange A.~M.,  Galland F.,  Udry S.,   Mayor M.,  2007, \mn@doi
  [A{\&}A] {10.1051/0004-6361:20078144}, 473, 983

\bibitem[\protect\citeauthoryear{Domingo et~al.,}{Domingo
  et~al.}{2009}]{Domingo2009}
Domingo V.,  et~al., 2009, Space Sci. Rev., 145, 337

\bibitem[\protect\citeauthoryear{Espinoza et~al.,}{Espinoza
  et~al.}{2019}]{Espinoza2019}
Espinoza N.,  et~al., 2019, \mn@doi [MNRAS] {10.1093/mnras/sty2691}, 482, 2065

\bibitem[\protect\citeauthoryear{Fligge, Solanki, Unruh, Frohlich  \&
  Wehrli}{Fligge et~al.}{1998}]{Fligge1998}
Fligge M.,  Solanki S.~K.,  Unruh Y.~C.,  Frohlich C.,   Wehrli C.,  1998,
  A{\&}A, 335, 709

\bibitem[\protect\citeauthoryear{Fligge, Solanki  \& Unruh}{Fligge
  et~al.}{2000}]{Fligge2000}
Fligge M.,  Solanki S.~K.,   Unruh Y.~C.,  2000, A{\&}A, 353, 380

\bibitem[\protect\citeauthoryear{Fr{\"{o}}hlich}{Fr{\"{o}}hlich}{2002}]{Frohlich2002}
Fr{\"{o}}hlich C.,  2002, \mn@doi [Advances in Space Research]
  {10.1016/S0273-1177(02)00203-X}, 29, 1409

\bibitem[\protect\citeauthoryear{Fr{\"{o}}hlich}{Fr{\"{o}}hlich}{2013}]{Frohlich2013}
Fr{\"{o}}hlich C.,  2013, \mn@doi [Space Sci. Rev.]
  {10.1007/s11214-011-9780-1}, 176, 237

\bibitem[\protect\citeauthoryear{Grossmann-Doerth, Schuessler  \&
  Steiner}{Grossmann-Doerth et~al.}{1998}]{GrossmannDoerth1998}
Grossmann-Doerth U.,  Schuessler M.,   Steiner O.,  1998, A{\&}A, 337, 928

\bibitem[\protect\citeauthoryear{Hale, Ellerman, Nicholson  \& Joy}{Hale
  et~al.}{1919}]{Hale1919}
Hale G.~E.,  Ellerman F.,  Nicholson S.~B.,   Joy A.~H.,  1919, \mn@doi [ApJ]
  {10.1086/142452}, 49, 153

\bibitem[\protect\citeauthoryear{Hall, Henry, Lockwood, Skiff  \& Saar}{Hall
  et~al.}{2009}]{Hall2009}
Hall J.~C.,  Henry G.~W.,  Lockwood G.~W.,  Skiff B.~A.,   Saar S.~H.,  2009,
  \mn@doi [AJ] {10.1088/0004-6256/138/1/312}, 138, 312

\bibitem[\protect\citeauthoryear{Haywood et~al.,}{Haywood
  et~al.}{2014}]{Haywood2014b}
Haywood R.~D.,  et~al., 2014, \mn@doi [MNRAS] {10.1093/mnras/stu1320}, 443,
  2517

\bibitem[\protect\citeauthoryear{Herrero, Ribas, Jordi, Morales, Perger  \&
  Rosich}{Herrero et~al.}{2016}]{Herrero2016}
Herrero E.,  Ribas I.,  Jordi C.,  Morales J.~C.,  Perger M.,   Rosich A.,
  2016, A{\&}A, 586, A131

\bibitem[\protect\citeauthoryear{Hirzberger et~al.,}{Hirzberger
  et~al.}{2010}]{Hirzberger2010}
Hirzberger J.,  et~al., 2010, \mn@doi [ApJ] {10.1088/2041-8205/723/2/L154},
  723, L154

\bibitem[\protect\citeauthoryear{I{\c{s}}ık, Solanki, Krivova  \&
  Shapiro}{I{\c{s}}ık et~al.}{2018}]{Isik2018}
I{\c{s}}ık E.,  Solanki S.~K.,  Krivova N.~A.,   Shapiro A.~I.,  2018, \mn@doi
  [A{\&}A] {10.1051/0004-6361/201833393}, 620, A177

\bibitem[\protect\citeauthoryear{I{\c{s}}ık, Shapiro, Solanki  \&
  Krivova}{I{\c{s}}ık et~al.}{2020}]{Isik2020}
I{\c{s}}ık E.,  Shapiro A.~I.,  Solanki S.~K.,   Krivova N.~A.,  2020, \mn@doi
  [ApJ] {10.3847/2041-8213/abb409}, 901, L12

\bibitem[\protect\citeauthoryear{Jackson \& Jeffries}{Jackson \&
  Jeffries}{2012}]{Jackson2012}
Jackson R.~J.,  Jeffries R.~D.,  2012, \mn@doi [MNRAS]
  {10.1111/j.1365-2966.2012.21119.x}, 423, 2966

\bibitem[\protect\citeauthoryear{Jackson \& Jeffries}{Jackson \&
  Jeffries}{2013}]{Jackson2013}
Jackson R.~J.,  Jeffries R.~D.,  2013, \mn@doi [MNRAS] {10.1093/mnras/stt304},
  431, 1883

\bibitem[\protect\citeauthoryear{Keller, Sch{\"{u}}ssler, V{\"{o}}gler  \&
  Zakharov}{Keller et~al.}{2004}]{Keller2004}
Keller C.~U.,  Sch{\"{u}}ssler M.,  V{\"{o}}gler A.,   Zakharov V.,  2004, ApJ,
  607, L59

\bibitem[\protect\citeauthoryear{Kirk, Wheatley, Louden, Littlefair,
  Copperwheat, Armstrong, Marsh  \& Dhillon}{Kirk et~al.}{2016}]{Kirk2016}
Kirk J.,  Wheatley P.~J.,  Louden T.,  Littlefair S.~P.,  Copperwheat C.~M.,
  Armstrong D.~J.,  Marsh T.~R.,   Dhillon V.~S.,  2016, \mn@doi [MNRAS]
  {10.1093/mnras/stw2205}, 463, 2922

\bibitem[\protect\citeauthoryear{Korhonen, Andersen, Piskunov, Hackman,
  Juncher, J{\"{a}}rvinen  \& J{\o}rgensen}{Korhonen
  et~al.}{2015}]{Korhonen2015}
Korhonen H.,  Andersen J.~M.,  Piskunov N.,  Hackman T.,  Juncher D.,
  J{\"{a}}rvinen S.~P.,   J{\o}rgensen U.~G.,  2015, \mn@doi [MNRAS]
  {10.1093/mnras/stu2730}, 448, 3038

\bibitem[\protect\citeauthoryear{Krivova, Solanki, Fligge  \& Unruh}{Krivova
  et~al.}{2003}]{Krivova2003}
Krivova N.~A.,  Solanki S.~K.,  Fligge M.,   Unruh Y.~C.,  2003, A{\&}A, 399,
  L1

\bibitem[\protect\citeauthoryear{Kurucz}{Kurucz}{2017}]{Kurucz2017}
Kurucz R.~L.,  2017, {ATLAS9: Model atmosphere program with opacity
  distribution functions}

\bibitem[\protect\citeauthoryear{Lagrange, Desort  \& Meunier}{Lagrange
  et~al.}{2010}]{Lagrange2010}
Lagrange A.~M.,  Desort M.,   Meunier N.,  2010, \mn@doi [A{\&}A]
  {10.1051/0004-6361/200913071}, 512, A38

\bibitem[\protect\citeauthoryear{Leighton}{Leighton}{1963}]{Leighton1963}
Leighton R.~B.,  1963, \mn@doi [ARA{\&}A]
  {10.1146/annurev.aa.01.090163.000315}, 1, 19

\bibitem[\protect\citeauthoryear{Lockwood, Skiff, Henry, Henry, Radick,
  Baliunas, Donahue  \& Soon}{Lockwood et~al.}{2007}]{Lockwood2007}
Lockwood G.~W.,  Skiff B.~A.,  Henry G.~W.,  Henry S.,  Radick R.~R.,  Baliunas
  S.~L.,  Donahue R.~A.,   Soon W.,  2007, \mn@doi [ApJS] {10.1086/516752},
  171, 260

\bibitem[\protect\citeauthoryear{Luger, Foreman-Mackey, Hedges  \& Hogg}{Luger
  et~al.}{2021}]{Rodrigo2021a}
Luger R.,  Foreman-Mackey D.,  Hedges C.,   Hogg D.~W.,  2021, arXiv e-prints,
  p. arXiv:2102.00007

\bibitem[\protect\citeauthoryear{Mancini et~al.,}{Mancini
  et~al.}{2014}]{Mancini2014}
Mancini L.,  et~al., 2014, \mn@doi [MNRAS] {10.1093/mnras/stu1286}, 443, 2391

\bibitem[\protect\citeauthoryear{Meunier \& Lagrange}{Meunier \&
  Lagrange}{2019}]{Meunier2019b}
Meunier N.,  Lagrange A.~M.,  2019, \mn@doi [A{\&}A]
  {10.1051/0004-6361/201935651}, 629, A42

\bibitem[\protect\citeauthoryear{Meunier, Desort  \& Lagrange}{Meunier
  et~al.}{2010a}]{Meunier2010a}
Meunier N.,  Desort M.,   Lagrange A.~M.,  2010a, \mn@doi [A{\&}A]
  {10.1051/0004-6361/200913551}, 512, A39

\bibitem[\protect\citeauthoryear{Meunier, Lagrange  \& Desort}{Meunier
  et~al.}{2010b}]{Meunier2010b}
Meunier N.,  Lagrange A.-M.,   Desort M.,  2010b, \mn@doi [A{\&}A]
  {10.1051/0004-6361/201014199}, 519, A66

\bibitem[\protect\citeauthoryear{Meunier, Lagrange, Boulet  \&
  Borgniet}{Meunier et~al.}{2019}]{Meunier2019a}
Meunier N.,  Lagrange A.~M.,  Boulet T.,   Borgniet S.,  2019, \mn@doi [A{\&}A]
  {10.1051/0004-6361/201834796}, 627, A56

\bibitem[\protect\citeauthoryear{Morris, Hebb, Davenport, Rohn  \&
  Hawley}{Morris et~al.}{2017}]{Morris2017}
Morris B.~M.,  Hebb L.,  Davenport J. R.~A.,  Rohn G.,   Hawley S.~L.,  2017,
  \mn@doi [ApJ] {10.3847/1538-4357/aa8555}, 846, 99

\bibitem[\protect\citeauthoryear{Neff, O'Neal  \& Saar}{Neff
  et~al.}{1995}]{Neff1995}
Neff J.~E.,  O'Neal D.,   Saar S.~H.,  1995, \mn@doi [ApJ] {10.1086/176356},
  452, 879

\bibitem[\protect\citeauthoryear{N{\`{e}}mec, Shapiro, Krivova, Solanki,
  Tagirov, Cameron  \& Dreizler}{N{\`{e}}mec et~al.}{2020a}]{Nemec2020a}
N{\`{e}}mec N.~E.,  Shapiro A.~I.,  Krivova N.~A.,  Solanki S.~K.,  Tagirov
  R.~V.,  Cameron R.~H.,   Dreizler S.,  2020a, \mn@doi [A{\&}A]
  {10.1051/0004-6361/202037588}, 636, A43

\bibitem[\protect\citeauthoryear{N{\`{e}}mec, I{\c{s}}ık, Shapiro, Solanki,
  Krivova  \& Unruh}{N{\`{e}}mec et~al.}{2020b}]{Nemec2020b}
N{\`{e}}mec N.~E.,  I{\c{s}}ık E.,  Shapiro A.~I.,  Solanki S.~K.,  Krivova
  N.~A.,   Unruh Y.,  2020b, \mn@doi [A{\&}A] {10.1051/0004-6361/202038054},
  638, A56

\bibitem[\protect\citeauthoryear{Norris}{Norris}{2018}]{Norris2018}
Norris C.~M.,  2018, PhD thesis, Imperial College London

\bibitem[\protect\citeauthoryear{Norris, Beeck, Unruh, Solanki, Krivova  \&
  Yeo}{Norris et~al.}{2017}]{Norris2017}
Norris C.~M.,  Beeck B.,  Unruh Y.~C.,  Solanki S.~K.,  Krivova N.~A.,   Yeo
  K.~L.,  2017, A{\&}A, 605, A45

\bibitem[\protect\citeauthoryear{O'Neal, Saar  \& Neff}{O'Neal
  et~al.}{1996}]{ONeal1996}
O'Neal D.,  Saar S.~H.,   Neff J.~E.,  1996, \mn@doi [ApJ] {10.1086/177288},
  463, 766

\bibitem[\protect\citeauthoryear{O'Neal, Neff, Saar  \& Cuntz}{O'Neal
  et~al.}{2004}]{ONeal2004}
O'Neal D.,  Neff J.~E.,  Saar S.~H.,   Cuntz M.,  2004, \mn@doi [AJ]
  {10.1086/423438}, 128, 1802

\bibitem[\protect\citeauthoryear{Oshagh, Santos, Ehrenreich, Haghighipour,
  Figueira, Santerne  \& Montalto}{Oshagh et~al.}{2014}]{Oshagh2014}
Oshagh M.,  Santos N.~C.,  Ehrenreich D.,  Haghighipour N.,  Figueira P.,
  Santerne A.,   Montalto M.,  2014, \mn@doi [A{\&}A]
  {10.1051/0004-6361/201424059}, 568, A99

\bibitem[\protect\citeauthoryear{Oshagh et~al.,}{Oshagh
  et~al.}{2017}]{Oshagh2017}
Oshagh M.,  et~al., 2017, \mn@doi [A{\&}A] {10.1051/0004-6361/201731139}, 606,
  A107

\bibitem[\protect\citeauthoryear{Panja, Cameron  \& Solanki}{Panja
  et~al.}{2020}]{Panja2020}
Panja M.,  Cameron R.,   Solanki S.~K.,  2020, \mn@doi [ApJ]
  {10.3847/1538-4357/ab8230}, 893, 113

\bibitem[\protect\citeauthoryear{Parker}{Parker}{1978}]{Parker1978}
Parker E.~N.,  1978, \mn@doi [ApJ] {10.1086/156035}, 221, 368

\bibitem[\protect\citeauthoryear{Rackham, Apai  \& Giampapa}{Rackham
  et~al.}{2018}]{Rackham2018}
Rackham B.~V.,  Apai D.,   Giampapa M.~S.,  2018, \mn@doi [ApJ]
  {10.3847/1538-4357/aaa08c}, 853, 122

\bibitem[\protect\citeauthoryear{Rackham, Apai  \& Giampapa}{Rackham
  et~al.}{2019}]{Rackham2019}
Rackham B.~V.,  Apai D.,   Giampapa M.~S.,  2019, \mn@doi [ApJ]
  {10.3847/1538-3881/aaf892}, 157, 96

\bibitem[\protect\citeauthoryear{Radick, Lockwood, Henry, Hall  \&
  Pevtsov}{Radick et~al.}{2018}]{Radick2018}
Radick R.~R.,  Lockwood G.~W.,  Henry G.~W.,  Hall J.~C.,   Pevtsov A.~A.,
  2018, ApJ, 855

\bibitem[\protect\citeauthoryear{Reinhold, Shapiro, Solanki, Montet, Krivova,
  Cameron  \& Amazo-G{\'{o}}mez}{Reinhold et~al.}{2020}]{Reinhold2020}
Reinhold T.,  Shapiro A.~I.,  Solanki S.~K.,  Montet B.~T.,  Krivova N.~A.,
  Cameron R.~H.,   Amazo-G{\'{o}}mez E.~M.,  2020, \mn@doi [Science]
  {10.1126/science.aay3821}, 368, 518

\bibitem[\protect\citeauthoryear{Rempel, Sch{\"{u}}ssler, Cameron  \&
  Kn{\"{o}}lker}{Rempel et~al.}{2009a}]{Rempel2009a}
Rempel M.,  Sch{\"{u}}ssler M.,  Cameron R.~H.,   Kn{\"{o}}lker M.,  2009a,
  \mn@doi [Science] {10.1126/science.1173798}, 325, 171

\bibitem[\protect\citeauthoryear{Rempel, Sch{\"{u}}ssler  \&
  Kn{\"{o}}lker}{Rempel et~al.}{2009b}]{Rempel2009b}
Rempel M.,  Sch{\"{u}}ssler M.,   Kn{\"{o}}lker M.,  2009b, \mn@doi [ApJ]
  {10.1088/0004-637X/691/1/640}, 691, 640

\bibitem[\protect\citeauthoryear{Ricker et~al.,}{Ricker
  et~al.}{2015}]{Ricker2015}
Ricker G.~R.,  et~al., 2015, \mn@doi [Journal of Astronomical Telescopes,
  Instruments, and Systems] {10.1117/1.JATIS.1.1.014003}, 1, 14003

\bibitem[\protect\citeauthoryear{Saar \& Donahue}{Saar \&
  Donahue}{1997}]{Saar1997}
Saar S.~H.,  Donahue R.~A.,  1997, \mn@doi [ApJ] {10.1086/304392}, 485, 319

\bibitem[\protect\citeauthoryear{Salabert, Garc{\'{i}}a, Jim{\'{e}}nez,
  Bertello, Corsaro  \& Pall{\'{e}}}{Salabert et~al.}{2017}]{Salabert2017}
Salabert D.,  Garc{\'{i}}a R.~A.,  Jim{\'{e}}nez A.,  Bertello L.,  Corsaro E.,
    Pall{\'{e}} P.~L.,  2017, \mn@doi [A{\&}A] {10.1051/0004-6361/201731560},
  608, A87

\bibitem[\protect\citeauthoryear{Salhab, Steiner, Berdyugina, Freytag, Rajaguru
   \& Steffen}{Salhab et~al.}{2018}]{Salhab2018}
Salhab R.~G.,  Steiner O.,  Berdyugina S.~V.,  Freytag B.,  Rajaguru S.~P.,
  Steffen M.,  2018, \mn@doi [A{\&}A] {10.1051/0004-6361/201731945}, 614, A78

\bibitem[\protect\citeauthoryear{Sarkar, Pascale, Papageorgiou, Johnson  \&
  Waldmann}{Sarkar et~al.}{2020}]{Sarkar2020}
Sarkar S.,  Pascale E.,  Papageorgiou A.,  Johnson L.~J.,   Waldmann I.,  2020,
  arXiv e-prints, p. arXiv:2002.03739

\bibitem[\protect\citeauthoryear{Scherrer et~al.,}{Scherrer
  et~al.}{1995}]{Scherrer1995}
Scherrer P.~H.,  et~al., 1995, \mn@doi [Solar Phys.] {10.1007/BF00733429}, 162,
  129

\bibitem[\protect\citeauthoryear{Schou et~al.,}{Schou et~al.}{2012}]{Schou2012}
Schou J.,  et~al., 2012, \mn@doi [Solar Phys.] {10.1007/s11207-011-9842-2},
  275, 229

\bibitem[\protect\citeauthoryear{Schrijver}{Schrijver}{2020}]{Schrijver2020}
Schrijver C.~J.,  2020, ApJ, 890, 121

\bibitem[\protect\citeauthoryear{Sch{\"{u}}ssler, Caligari, Ferriz-Mas, Solanki
   \& Stix}{Sch{\"{u}}ssler et~al.}{1996}]{Schussler1996}
Sch{\"{u}}ssler M.,  Caligari P.,  Ferriz-Mas A.,  Solanki S.~K.,   Stix M.,
  1996, A{\&}A, 314, 503

\bibitem[\protect\citeauthoryear{Shapiro, Solanki, Krivova, Schmutz, Ball,
  Knaack, Rozanov  \& Unruh}{Shapiro et~al.}{2014}]{Shapiro2014}
Shapiro A.~I.,  Solanki S.~K.,  Krivova N.~A.,  Schmutz W.~K.,  Ball W.~T.,
  Knaack R.,  Rozanov E.~V.,   Unruh Y.~C.,  2014, A{\&}A, 569

\bibitem[\protect\citeauthoryear{Shapiro, Solanki, Krivova, Cameron, Yeo  \&
  Schmutz}{Shapiro et~al.}{2017}]{Shapiro2017}
Shapiro A.~I.,  Solanki S.~K.,  Krivova N.~A.,  Cameron R.~H.,  Yeo K.~L.,
  Schmutz W.~K.,  2017, \mn@doi [Nature Astronomy] {10.1038/s41550-017-0217-y},
  1, 612

\bibitem[\protect\citeauthoryear{Shapiro, Amazo-G{\'{o}}mez, Krivova  \&
  Solanki}{Shapiro et~al.}{2020}]{Shapiro2020}
Shapiro A.~I.,  Amazo-G{\'{o}}mez E.~M.,  Krivova N.~A.,   Solanki S.~K.,
  2020, \mn@doi [A{\&}A] {10.1051/0004-6361/201936018}, 633, A32

\bibitem[\protect\citeauthoryear{Silva-Valio, Lanza, Alonso  \&
  Barge}{Silva-Valio et~al.}{2010}]{Silva-Valio2010b}
Silva-Valio A.,  Lanza A.~F.,  Alonso R.,   Barge P.,  2010, \mn@doi [A{\&}A]
  {10.1051/0004-6361/200911904}, 510, A25

\bibitem[\protect\citeauthoryear{Sing}{Sing}{2010}]{Sing2010}
Sing D.~K.,  2010, \mn@doi [A{\&}A] {10.1051/0004-6361/200913675}, 510, A21

\bibitem[\protect\citeauthoryear{Sing, D{\'{e}}sert, Lecavelier Des~Etangs,
  Ballester, Vidal-Madjar, Parmentier, Hebrard  \& Henry}{Sing
  et~al.}{2009}]{Sing2009}
Sing D.~K.,  D{\'{e}}sert J.~M.,  Lecavelier Des~Etangs A.,  Ballester G.~E.,
  Vidal-Madjar A.,  Parmentier V.,  Hebrard G.,   Henry G.~W.,  2009, \mn@doi
  [A{\&}A] {10.1051/0004-6361/200912776}, 505, 891

\bibitem[\protect\citeauthoryear{Solanki}{Solanki}{1993}]{Solanki1993}
Solanki S.~K.,  1993, \mn@doi [Space Sci. Rev.] {10.1007/BF00749277}, 63, 1

\bibitem[\protect\citeauthoryear{Solanki}{Solanki}{2003}]{Solanki2003}
Solanki S.~K.,  2003, \mn@doi [A{\&}ARv] {10.1007/s00159-003-0018-4}, 11, 153

\bibitem[\protect\citeauthoryear{Solanki \& Unruh}{Solanki \&
  Unruh}{2004}]{Solanki2004}
Solanki S.~K.,  Unruh Y.~C.,  2004, MNRAS, 348, 307

\bibitem[\protect\citeauthoryear{Solanki \& Unruh}{Solanki \&
  Unruh}{2013}]{Solanki2013}
Solanki S.~K.,  Unruh Y.~C.,  2013, Astron. Nachr., 334, 145

\bibitem[\protect\citeauthoryear{Spruit}{Spruit}{1976}]{Spruit1976}
Spruit H.~C.,  1976, \mn@doi [Solar Phys.] {10.1007/BF00155292}, 50, 269

\bibitem[\protect\citeauthoryear{Spruit}{Spruit}{1979}]{Spruit1979}
Spruit H.~C.,  1979, \mn@doi [Solar Phys.] {10.1007/BF00150420}, 61, 363

\bibitem[\protect\citeauthoryear{Strassmeier}{Strassmeier}{2009}]{Strassmeier2009}
Strassmeier K.~G.,  2009, \mn@doi [A{\&}ARv] {10.1007/s00159-009-0020-6}, 17,
  251

\bibitem[\protect\citeauthoryear{Torres, Ferraz~Mello  \& Quast}{Torres
  et~al.}{1972}]{Torres1972}
Torres C.~A.~O.,  Ferraz~Mello S.,   Quast G.~R.,  1972, Astrophys. Lett., 11,
  13

\bibitem[\protect\citeauthoryear{Unruh, Solanki  \& Fligge}{Unruh
  et~al.}{1999}]{Unruh1999}
Unruh Y.~C.,  Solanki S.~K.,   Fligge M.,  1999, A{\&}A, 345, 635

\bibitem[\protect\citeauthoryear{Unruh, Krivova, Solanki, Harder  \&
  Kopp}{Unruh et~al.}{2008}]{Unruh2008}
Unruh Y.~C.,  Krivova N.~A.,  Solanki S.~K.,  Harder J.~W.,   Kopp G.,  2008,
  A{\&}A, 486, 311

\bibitem[\protect\citeauthoryear{V{\"{o}}gler, Shelyag, Sch{\"{u}}ssler,
  Cattaneo, Emonet  \& Linde}{V{\"{o}}gler et~al.}{2005}]{Vogler2005}
V{\"{o}}gler A.,  Shelyag S.,  Sch{\"{u}}ssler M.,  Cattaneo F.,  Emonet T.,
  Linde T.,  2005, A{\&}A, 429, 335

\bibitem[\protect\citeauthoryear{Wenzler, Solanki, Krivova  \&
  Fr{\"{o}}hlich}{Wenzler et~al.}{2006}]{Wenzler2006}
Wenzler T.,  Solanki S.~K.,  Krivova N.~A.,   Fr{\"{o}}hlich C.,  2006, \mn@doi
  [A{\&}A] {10.1051/0004-6361:20065752}, 460, 583

\bibitem[\protect\citeauthoryear{Willson \& Hudson}{Willson \&
  Hudson}{1991}]{Wilson1991}
Willson R.~C.,  Hudson H.~S.,  1991, \mn@doi [Nature] {10.1038/351042a0}, 351,
  42

\bibitem[\protect\citeauthoryear{Wolter, Schmitt  \& van Wyk}{Wolter
  et~al.}{2005}]{Wolter2005}
Wolter U.,  Schmitt J.~H.~M.~M.,   van Wyk F.,  2005, \mn@doi [A{\&}A]
  {10.1051/0004-6361:20042239}, 435, 261

\bibitem[\protect\citeauthoryear{Wolter, Robrade, Schmitt  \& Ness}{Wolter
  et~al.}{2008}]{Wolter2008}
Wolter U.,  Robrade J.,  Schmitt J.~H.~M.~M.,   Ness J.~U.,  2008, \mn@doi
  [A{\&}A] {10.1051/0004-6361:20078838}, 478, L11

\bibitem[\protect\citeauthoryear{Wolter, Schmitt, Huber, Czesla, M{\"{u}}ller,
  Guenther  \& Hatzes}{Wolter et~al.}{2009}]{Wolter2009}
Wolter U.,  Schmitt J.~H.~M.~M.,  Huber K.~F.,  Czesla S.,  M{\"{u}}ller H.~M.,
   Guenther E.~W.,   Hatzes A.~P.,  2009, \mn@doi [A{\&}A]
  {10.1051/0004-6361/200912329}, 504, 561

\bibitem[\protect\citeauthoryear{Yeo, Solanki  \& Krivova}{Yeo
  et~al.}{2013}]{Yeo2013}
Yeo K.~L.,  Solanki S.~K.,   Krivova N.~A.,  2013, \mn@doi [A{\&}A]
  {10.1051/0004-6361/201220682}, 550, A95

\bibitem[\protect\citeauthoryear{Yeo, Krivova, Solanki  \& Glassmeier}{Yeo
  et~al.}{2014}]{Yeo2014}
Yeo K.~L.,  Krivova N.~A.,  Solanki S.~K.,   Glassmeier K.~H.,  2014, \mn@doi
  [A{\&}A] {10.1051/0004-6361/201423628}, 570, A85

\bibitem[\protect\citeauthoryear{Yeo, Solanki, Norris, Beeck, Unruh  \&
  Krivova}{Yeo et~al.}{2017}]{Yeo2017}
Yeo K.~L.,  Solanki S.~K.,  Norris C.~M.,  Beeck B.,  Unruh Y.~C.,   Krivova
  N.~A.,  2017, Physical Review Letters, 119

\bibitem[\protect\citeauthoryear{van Leeuwen, Alphenaar  \& Meys}{van Leeuwen
  et~al.}{1987}]{VanLeeuwen1987}
van Leeuwen F.,  Alphenaar P.,   Meys J.~J.~M.,  1987, A{\&}AS, 67, 483

\makeatother
\end{thebibliography}

\begin{appendix}

\section{Robustness of $R_{\rm var}$} \label{noise}

\begin{figure}
\centering
\includegraphics[width=\hsize]{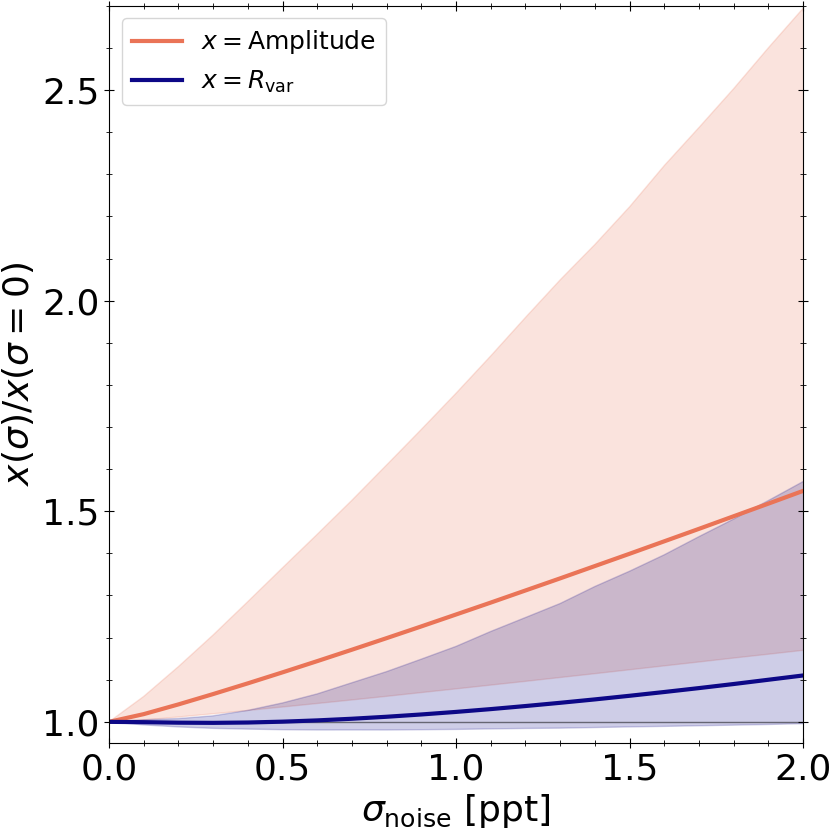}
\caption{\label{fig:noisesims}Plot of fractional change in variability proxy against noise level for both the untrimmed lightcurve amplitude (red) and $R_{\rm var}$ (blue). Filled lines represent the means and shaded regions represent the 95\% confidence intervals.}
\end{figure}

While we do not include instrumental noise in our simulations, we evaluated the robustness of $R_{\rm var}$ compared to the untrimmed lightcurve amplitude by using a Gaussian noise generator to add instrumental noise to a set of \texttt{actress} lightcurves. The fractional change in both variability proxies is plotted against noise level in Fig.~\ref{fig:noisesims}.

The presence of noise typically increases both $R_{\rm var}$ and the amplitude. The mean increase for $R_{\rm var}$ at a noise level of $1\,{\rm ppt}$ (roughly comparable to the estimated precision for long-cadence observations of a \textit{Kepler} magnitude $K_{\rm p} = 15.5$ target) is of the order of 2\%, while the corresponding mean amplitude increase is roughly 25\%. However, the range of changes in $R_{\rm var}$ and amplitude are large. At the 95\% confidence level, the range in $R_{\rm var}$ change is $[-4\%,15\%]$ and the range in amplitude change is $[-17\%,53\%]$. These results confirm the relative robustness of $R_{\rm var}$ to noise.

\section{\textit{TESS}-band intensity coefficients and quadratic equivalents} \label{app:tessfit}

We provide limb-dependent intensity coefficients for the quiet photosphere and facular regions in the \textit{TESS} band (Table \ref{table:LDTess}), for all available spectral types and magnetic field strengths. Variability results calculated in the \textit{TESS} band are presented and compared with \textit{Kepler}-band variability in Sect.~\ref{tess}. In addition, we provide equivalent \textit{Kepler} and \textit{TESS}-band fit coefficients for the quadratic limb darkening law (Tables \ref{table:LDquad} and \ref{table:LDquadTESS})

\begin{equation}
I(\mu) = I(1) \left( 1 - a(1-\mu) -  b(1-\mu^2) \right),
\label{eq:limb_darkening_quad}
\end{equation}
where $I(1)$ is the intensity at disc centre and $a$ and $b$ are limb darkening coefficients.

\begin{table}
\caption{Limb-dependent \textit{TESS}-band intensity coefficients for the 3-parameter non-linear limb darkening law (Equation \ref{eq:limb_darkening}), for all available spectral types and magnetic field strengths in our modelling approach.}      
\label{table:LDTess} 
\centering             
\begin{tabular}{c c c c c c}  
\hline\hline                 
Type  & $\langle B_{\rm z} \rangle$ & $I(1)$ & $a$ & $b$ & $c$ \cr & [G] & [$\rm photons/m^{2}/s/sr$] & & &\\  

\hline                        
      &  hydro & $3.12 \times 10^{21}$ & 1.48 & -1.29 & 0.43\\
  G2  &  100 & $3.16 \times 10^{21}$ & 1.55 & -1.43 & 0.48\\
      &  500 & $3.13 \times 10^{21}$ & 2.00 & -2.37 & 0.92\\
      \hline
      &  hydro & $1.70 \times 10^{21}$ & 1.05 & -0.46 & 0.07\\
  K0  &  100 & $1.72 \times 10^{21}$ & 1.17 & -0.70 & 0.18\\
      &  500 & $1.71 \times 10^{21}$ & 1.30 & -1.00 & 0.32\\
      \hline
      &  hydro & $5.86 \times 10^{20}$ & 1.82 & -1.79 & 0.63\\
  M0  &  100 & $5.87 \times 10^{20}$ & 1.82 & -1.81 & 0.63\\
      &  500 & $5.79 \times 10^{20}$ & 1.73 & -1.70  & 0.59\\
      \hline
      &  hydro & $4.31 \times 10^{20}$ & 1.80 & -1.76 & 0.60\\
  M2  &  100 & $4.31 \times 10^{20}$ & 1.68 & -1.54 & 0.48\\
      &  500 & $4.21 \times 10^{20}$ & 1.96 & -2.16 & 0.82\\
\hline                                   
\end{tabular}
\end{table}

\begin{table}
\caption{Limb-dependent \textit{Kepler}-band intensity coefficients for the quadratic law (Equation \ref{eq:limb_darkening_quad}), for all available spectral types and magnetic field strengths in our modelling approach.}      
\label{table:LDquad} 
\centering             
\begin{tabular}{c c c c c}  
\hline\hline                 
Type  & $\langle B_{\rm z} \rangle$ & $I(1)$ & $a$ & $b$ \cr & [G] & [$\rm photons/m^{2}/s/sr$] & &\\  

\hline                        
      &  hydro & $2.33 \times 10^{21}$ & 0.86 & -0.21\\
  G2  &  100 & $2.36 \times 10^{21}$ & 0.86 & -0.22\\
      &  500 & $2.32 \times 10^{21}$ & 0.83 & -0.30\\
      \hline
      &  hydro & $1.10 \times 10^{21}$ & 0.85 & -0.13\\
  K0  &  100 & $1.12 \times 10^{21}$ & 0.85 & -0.14\\
      &  500 & $1.10 \times 10^{21}$ & 0.83 & -0.17\\
      \hline
      &  hydro & $2.79 \times 10^{20}$ & 0.95 & -0.27\\
  M0  &  100 & $2.80 \times 10^{20}$ & 0.95 & -0.28\\
      &  500 & $2.76 \times 10^{20}$ & 0.91 & -0.27\\
      \hline
      &  hydro & $1.85 \times 10^{20}$ & 0.94 & -0.27\\
  M2  &  100 & $1.86 \times 10^{20}$ & 0.94 & -0.28\\
      &  500 & $1.80 \times 10^{20}$ & 0.88 & -0.25\\
\hline                                   
\end{tabular}
\end{table}

\begin{table}
\caption{Limb-dependent \textit{TESS}-band intensity coefficients for the quadratic law (Equation \ref{eq:limb_darkening_quad}), for all available spectral types and magnetic field strengths in our modelling approach.}      
\label{table:LDquadTESS} 
\centering             
\begin{tabular}{c c c c c}  
\hline\hline                 
Type  & $\langle B_{\rm z} \rangle$ & $I(1)$ & $a$ & $b$ \cr & [G] & [$\rm photons/m^{2}/s/sr$] & &\\  

\hline                        
      &  hydro & $3.11 \times 10^{21}$ & 0.78 & -0.22\\
  G2  &  100 & $3.15 \times 10^{21}$ & 0.77 & -0.23\\
      &  500 & $3.11 \times 10^{21}$ & 0.72 & -0.27\\
      \hline
      &  hydro & $1.70 \times 10^{21}$ & 0.80 & -0.17\\
  K0  &  100 & $1.72 \times 10^{21}$ & 0.79 & -0.17\\
      &  500 & $1.70 \times 10^{21}$ & 0.76 & -0.18\\
      \hline
      &  hydro & $5.84 \times 10^{20}$ & 0.84 & -0.27\\
  M0  &  100 & $5.85 \times 10^{20}$ & 0.84 & -0.27\\
      &  500 & $5.77 \times 10^{20}$ & 0.80 & -0.27\\
      \hline
      &  hydro & $4.29 \times 10^{20}$ & 0.84 & -0.28\\
  M2  &  100 & $4.30 \times 10^{20}$ & 0.84 & -0.29\\
      &  500 & $4.19 \times 10^{20}$ & 0.78 & -0.26\\
\hline                                   
\end{tabular}
\end{table}

\section{Spot temperature determination from observations} \label{app:Tspot}

\begin{figure*}
\centering
\includegraphics[width=\hsize]{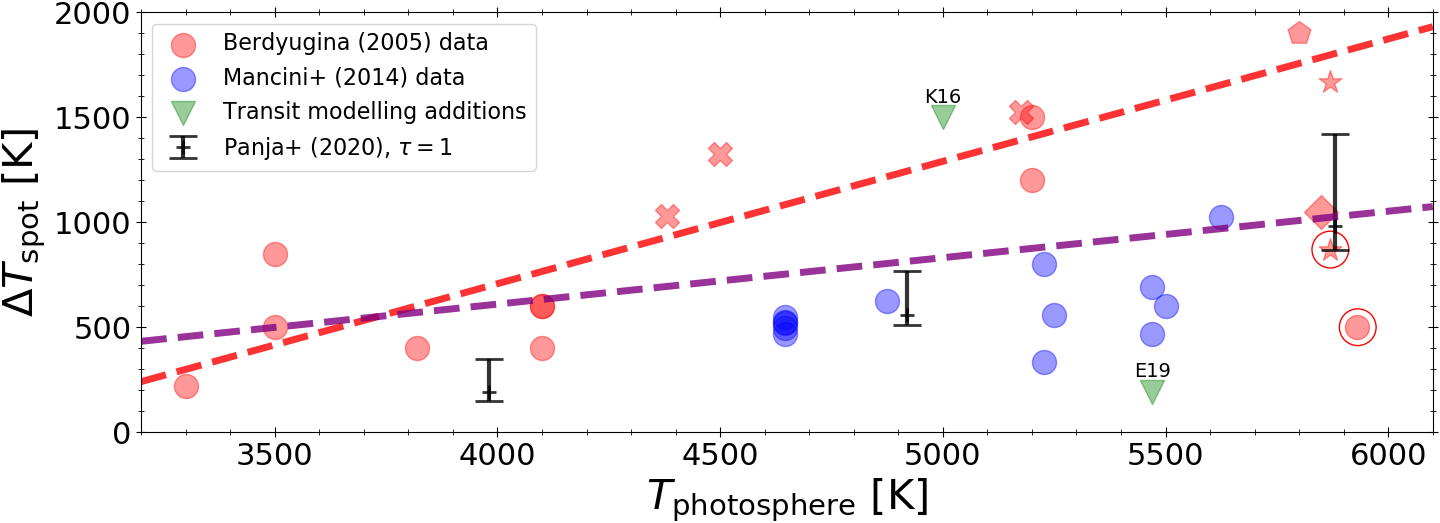}
\caption{\label{fig:TspotRel}Spot temperature difference, $\Delta T_{\rm spot} (= T_{\rm phot} - T_{\rm spot})$ against photospheric temperature, $T_{\rm phot}$ for dwarf stars collated in \citet{Berdyugina2005} (red), \citet{Mancini2014} (blue), with additions (in green) from K16 \citep{Kirk2016} and E19 \citep{Espinoza2019}. The red, dashed line represents the linear fit from \citet{Rackham2019} to the \citet{Berdyugina2005} data, and the purple, dashed line represents an unweighted linear fit made to the combined data. Spot temperatures were measured through lightcurve analysis (filled circles), TiO-band modelling (crosses), Doppler imaging (diamonds), transit modelling (triangles) and the combination of Doppler imaging and lightcurve analysis (pentagon). The two `star' markers represent sunspot umbra (upper) and penumbra (lower). Red circles around markers represent data excluded in the \citet{Rackham2019} fit. Black bars represent the umbral, average and penumbral temperatures (top to bottom) at optical depth $\tau = 1$ for MURaM-simulated starspots \citep{Panja2020}.}
\end{figure*}

Fig.~\ref{fig:TspotRel} shows a collection of spot temperature measurements from the literature. In principle, these can be used to re-determine the spot temperature dependence on spectral type. Spot temperatures collected here were determined using several different observational methods (lightcurve analysis, TiO-band modelling, Doppler imaging, line-depth ratios and transit modelling). These data were collected by \citet{Berdyugina2005} and \citet{Mancini2014}, with individual additions from \citet{ONeal2004}, \citet{Kirk2016} and \citet{Espinoza2019}. An unweighted linear fit to all available data is shown, alongside the \citet{Rackham2019} fit to the \citet{Berdyugina2005} data and the umbral, penumbral and average (assuming a 4:1 penumbral-to-umbral ratio) spot temperatures at optical depth $\tau=1$ derived from MURaM spot simulations \citep{Panja2020}.

\begin{figure}
\centering
\includegraphics[width=\hsize]{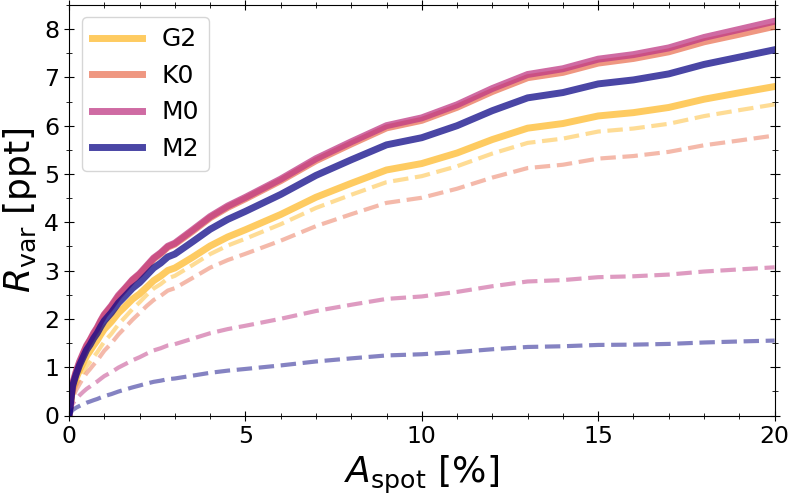}
\caption{\label{fig:TspotOld}Plot of mean range variability $R_{\rm var}$ against spot area coverage $A_{\rm spot}$ for all spectral types, with observationally-derived spot temperature contrasts (from the fit to all data in Fig.~\ref{fig:TspotRel}) at stellar inclination $i=90^\circ$ and weak-field facular regions. Variability results using \citet{Panja2020} spot contrasts are plotted for comparison (thin, dashed lines).}
\end{figure}

In Fig.~\ref{fig:TspotOld}, plots of mean $R_{\rm var}$ calculated with spot temperatures from the linear fit to the combined data are shown alongside the variability results using \citet{Panja2020} spot contrasts, for all spectral types with weak-field facular regions. Variability levels are higher for all spectral types due to the larger spot temperature differences from the observationally derived fit, and the spectral type dependence is lost due to the greater discrepancy between the fit and the \citet{Panja2020} spot contrasts for later spectral types. This proves spot temperature to be a critical parameter for visible-band variability, further highlighting the importance of its accurate determination for stellar lightcurve modelling.

\section{Feature distribution vs. activity level} \label{app:featdist}

\begin{figure}
\centering
\includegraphics[width=\hsize]{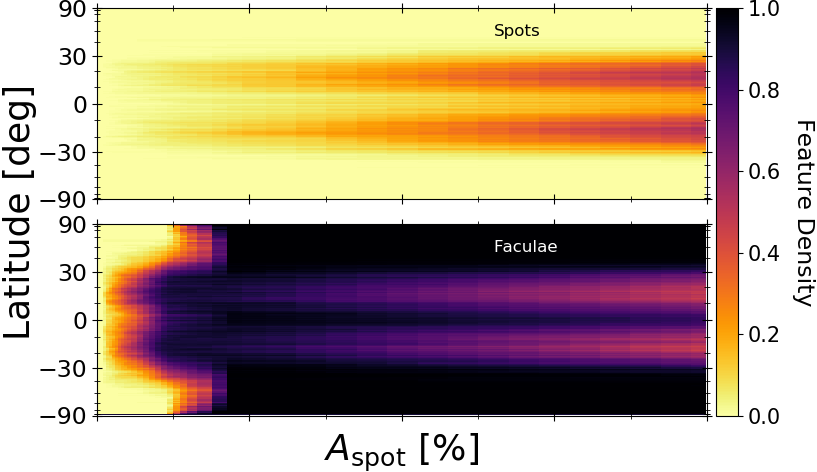}
\caption{\label{fig:fds}Spot and facular region latitude distributions against spot coverage for a single Monte-Carlo simulation run. A feature density of $1.0$ represents total coverage.}
\end{figure}

Fig.~\ref{fig:fds} shows how the latitude distributions of spots and facular regions change with spot coverage (or activity level) for the Monte-Carlo simulations described in Sect.~\ref{mcp}. At low coverage levels ($A_{\rm spot} \lessapprox 1.9\%$ or $A_{\rm tot} < 50\%$), the feature distributions are governed by the probability distributions shown in Fig.~\ref{fig:Dists} (top panel), with spot and facular bands centred on $\pm 16^\circ$. The facular bands are broader than the spot bands.

As coverage level increases beyond this point, the latitude constraint on faculae is removed, allowing high-latitude placement with a uniform distribution. The regions nearer the poles are populated before the mid-latitude ranges as there is less available surface area at higher latitudes. At $A_{\rm spot} \gtrapprox 3.8\%$, the stellar surface is completely covered with spots and facular regions and the facular-to-spot coverage fraction decreases as more spots are added. We note that this appears to occur at a higher spot coverage in Fig.~\ref{fig:fds} due to the binning. The bin centres correspond to the sampling intervals for the lightcurve simulations.

\end{appendix}
\label{lastpage}
\end{document}